\documentstyle[aps,prb,epsf,twocolumn]{revtex}

\catcode`\@=11
\def\simle{\mathrel{\mathpalette\@versim<}}   
\def\simge{\mathrel{\mathpalette\@versim>}}   
\def\@versim#1#2{\lower2.5pt\vbox{\baselineskip0pt \lineskip-.5pt
   \ialign{$\m@th#1\hfil##\hfil$\crcr#2\crcr\sim\crcr}}}

\begin{document}

\draft
\twocolumn[\hsize\textwidth\columnwidth\hsize\csname @twocolumnfalse\endcsname

\title{
Ordering and Fluctuation of Orbital and Lattice Distortion\\
in Perovskite Manganese Oxides
}
\author{Yukitoshi Motome}
\address{
Department of Physics, Tokyo Institute of Technology,\\
Oh-okayama 2-12-1, Meguro-ku, Tokyo 152-8551
}
\author{Masatoshi Imada}
\address{
Institute for Solid State Physics, University of Tokyo,\\
Roppongi 7-22-1, Minato-ku, Tokyo 106-8666
}
\date{
\today
}
\maketitle

\begin{abstract}
Roles of orbital and lattice degrees of freedom
in strongly correlated systems are investigated
to understand electronic properties of perovskite Mn oxides
such as La$_{1-x}$Sr$_{x}$MnO$_{3}$.
An extended double-exchange model containing Coulomb interaction,
doubly degenerate orbitals and Jahn-Teller coupling
is derived under full polarization of spins
with two-dimensional anisotropy.
Quantum fluctuation effects of Coulomb interaction and
orbital degrees of freedom are investigated
by using the quantum Monte Carlo method.
In undoped states,
it is crucial to consider both the Coulomb interaction
and the Jahn-Teller coupling
in reproducing characteristic hierarchy of energy scales
among charge, orbital-lattice and spin degrees of freedom
in experiments.
Our numerical results quantitatively reproduce
the charge gap amplitude as well as
the stabilization energy and the amplitude of
the cooperative Jahn-Teller distortion in undoped compounds.
Upon doping of carriers,
in the absence of the Jahn-Teller distortion,
critical enhancement of 
both charge compressibility and orbital correlation length 
is found with decreasing doping concentration.
These are discussed as origins of
strong incoherence in charge dynamics.
With the Jahn-Teller coupling in the doped region,
collapse of the Jahn-Teller distortion and instability
to phase separation are obtained and 
favorably compared with experiments.
These provide a possible way to understand
the complicated properties of lightly doped manganites.
\end{abstract}

\pacs{PACS numbers: 71.30.+h, 71.27.+a, 71.20.Be, 71.38.+i}
]

\section{Introduction}
\label{Sec:Introduction}

Perovskite Mn oxides show a variety of interesting physics
depending on chemical composition and temperature.
Pioneering work in 1950's revealed fascinating phase transitions
from an antiferromagnetic insulator to a ferromagnetic metal.
\cite{Jonker1950,Santen1950,Zener1951,Wollan1955,Goodenough1955,Anderson1955,deGennes1960}
The transitions take place through
a change in the chemical composition which effectively
introduces mobile holes in the systems.
In 1970, a large negative magnetoresistance was discovered
near the transition from the ferromagnetic metal to
a paramagnetic state with increasing temperature.
\cite{Searle1970,Searle1969}
Recently, this so-called colossal magnetoresistance
has attracted renewed interest
not only for understanding basic mechanisms but also
for a possibility of technological applications.
\cite{Ramirez1997}

The basic physics in these systems has been examined
by the so-called double-exchange mechanism.
\cite{Zener1951}
In undoped states, each Mn$^{3+}$ ion has four $3d$ electrons
in the $(t_{2g})^{3} (e_{g})^{1}$ configuration.
Doped mobile holes are introduced to the twofold $e_{g}$ orbitals.
These situations have been examined based on
the simplest version of the double-exchange (DE) model.
\cite{Zener1951,Anderson1955,deGennes1960,Kubo1972,Furukawa1994,Furukawa1998}
The model consists of
noninteracting conduction electrons in a nondegenerate orbital
and localized spins ferromagnetically coupled to this conduction band
through the Hund's rule coupling.
\cite{Zener1951}
The explicit form of the Hamiltonian is given by
\begin{equation}
        \label{H DE}
        {\cal H}_{\rm DE} = \sum_{ij}\sum_{\sigma}
        t_{ij} c_{i\sigma}^{\dagger}c_{j\sigma}
        + K \sum_{i} \mbox{\boldmath $\sigma$}_{i} \cdot 
        \mbox{\boldmath $S$}_{i},
\end{equation}
where $c_{i\sigma}^{\dagger} (c_{i\sigma})$ creates (annihilates)
a $\sigma$-spin electron at site $i$;
$\mbox{\boldmath $\sigma$}_{i}$ is the spin operator for conduction electrons
whereas $\mbox{\boldmath $S$}_{i}$ denotes the localized spin at site $i$.
This model successfully explains
the metallic ferromagnetism under hole doping,
the transition to paramagnetic phase with increasing temperature
and the colossal magnetoresistance.

Recently, intensive experimental studies have revealed
an importance of interplay among
charge, spin, orbital and lattice degrees of freedom in the Mn oxides.
\cite{Ramirez1997}
These systems show a Mott insulating or a charge-ordered state
concomitant with spin, orbital and lattice orderings;
a metallic state with different spin and orbital orderings;
and various transitions between them.
These novel phenomena strongly indicate a necessity
for studies beyond the `simple' DE model given by Eq. (\ref{H DE}).

In particular, we discuss here the following experimental results
which are difficult to understand within the `simple' DE model:
In undoped states, these materials are insulating
with a charge gap of the order of $1$eV.
At low temperatures, this state shows spin, orbital and lattice orderings
with strong two-dimensional (2D) anisotropy.
The magnetic structure is called $A$-type
and consists of ferromagnetic planes stacked antiferromagnetically.
For instance, in LaMnO$_{3}$, the charge gap amplitude $\Delta_{\rm c}$
is estimated around $1$eV,
\cite{Bocquet1992,Chainani1993,Arima1993,Saitoh1995}
the transition to the 2D orbital-ordered state
with lattice distortion (cooperative Jahn-Teller (JT) distortion)
occurs at $T_{\rm c} \simeq 780$K,
\cite{Wollan1955,Goodenough1955,Matsumoto1970a,Matsumoto1970b,Ellemans1971,Murakami1998}
and the $A$-type antiferromagnetic phase appears at $T_{\rm N} \simeq
140$K.
\cite{Wollan1955,Goodenough1955,Matsumoto1970a}
These indicate a hierarchy structure of energy scales:
$\Delta_{\rm c} \gg T_{\rm c} \gg T_{\rm N}$.

When we dope holes, the materials show complex transitions
at low temperatures;
collapse of the JT distortion, 
magnetic and metal-insulator (MI) transitions.
\cite{Jonker1950,Santen1950,Wollan1955,Goodenough1955,Matsumoto1970b,Tokura1994,Urushibara1995,Kawano1996a,Kawano1996b,Yamada1996,Zhou1997}
Doped holes generally induce a spin canting;
the spin direction in the 2D ferromagnetic planes tilts 
and a net ferromagnetic moment is
induced without destroying the cooperative JT distortion.
\cite{Wollan1955,Goodenough1955,Matsumoto1970b,Kawano1996a,Kawano1996b}
Recently, neutron scattering experiment reported
a charge or polaron ordering
with 2D anisotropy around $\delta = 0.125$
in the case of La$_{1-x}$Sr$_{x}$MnO$_{3}$
(Sr concentration $x$ corresponds to the hole concentration $\delta$).
\cite{Yamada1996,Zhou1997}
In this phase, the cooperative JT distortion is not observed
whereas the spins are canted.
The distortion collapses at $\delta \simeq 0.1$.
The system remains insulating through the spin-canted state
and finally at $\delta \simeq 0.16$,
becomes the ferromagnetic metal.

An important and puzzling feature appears in
the ferromagnetic state near the MI transition;
in the case of La$_{1-x}$Sr$_{x}$MnO$_{3}$, for $\delta \simge 0.16$.
The metal exhibits strong charge incoherence at low temperatures
where spin degrees of freedom are frozen up
due to the perfect polarization.
A strong mass enhancement is suggested by
a small Drude weight $D$ in the optical conductivity
\cite{Okimoto1995,Okimoto1997}
and 
a small discontinuity at the Fermi edge in the photoemission spectra,
\cite{Sarma1996}
while there is no conspicuous enhancement in
the specific heat coefficient $\gamma$.
\cite{Woodfield1997}

In order to consider the above experimental aspects,
several approaches beyond the `simple' DE model (\ref{H DE})
have been explored.
One is mean-field approximation.
The undoped states have been investigated
by effective Hamiltonian
derived from a strong coupling limit of electron correlations,
\cite{Kugel1973,Shiina1997,Ishihara1997a,Feiner1999}
by extended DE models with both spin and orbital degrees of
freedom,
\cite{Maezono1998}
and by models containing oxygens.
\cite{Mizokawa1995,Mizokawa1996}
These mean-field results commonly suggest energetic degeneracy
between different spin and orbital structures.
The importance of JT coupling or
$t_{2g}$-spin interaction has been discussed
to lift this degeneracy and to reproduce the realistic state.
The following points, however, are missing in these calculations:
They have not seriously considered
the experimental energetic hierarchy,
$\Delta_{\rm c} \gg T_{\rm c} \gg T_{\rm N}$.
Although the JT distortion appears in much higher energy scale
than the spin ordering temperatures,
the optimization of the ground state including phonons
has not been successfully done.
The model parameters, for instance, the interaction strength or
the JT coupling they have chosen,
are rather scattered and the consensus of parameter values
is not reached.
Among others, it should be noted that,
as suggested in some degeneracy of various phases in these calculations,
the magnetic and orbital structure
must be under strong quantum fluctuations.

There is another approach, first-principles calculation,
which is applied mainly to undoped states.
The local-density approximation (LDA) and
the local-spin-density approximation (LSDA) conclude
a gapless metal for the undoped state
when the JT distortions are neglected.
On the assumption of the experimental lattice structure,
LSDA gives the antiferromagnetic insulator,
although the charge gap amplitude is insufficient
in reproducing the experimental result.
\cite{Satpathy1996,Solovyev1996a}
These are a consequence of the general tendency
to underestimate electron correlation effects
in these approximations.
LDA+U calculations, in which
the effects of strong correlation are taken into account
by a Hartree-Fock scheme, have also been done
under the experimental lattice structure and
reproduced realistic values of the charge gap amplitude
and the magnetic moment.
\cite{Satpathy1996,Solovyev1996b}
The improvement reached in LDA+U with the JT distortion implies that
not only the JT coupling but also the strong correlation
play an important role in these materials.
The correct description of the experimentally-observed ground state,
especially the large charge-gap amplitude,
has not been reached ab-initio.
We note that it is rather hard to examine
quantum fluctuation effects by the available
band-structure calculations.

In contrast to the undoped states, doping effects have been
less investigated theoretically.
Especially, the JT distortions in doped regions
have not been seriously examined.
A systematic study on the doping effects is
desired to discuss the collapse of the distortions and
the charge or polaron ordering in experiments.

In the ferromagnetic metal,
the strong charge incoherence has been discussed
beyond the `simple' DE models.
Since the spins are perfectly polarized at low temperatures,
this incoherence strongly suggests
an importance of neglected elements in the model.
Many candidates of the driving mechanism have been discussed
such as JT fluctuation,
\cite{Millis1996a}
interorbital hopping of the $e_{g}$ orbitals,
\cite{Shiba1997,Brito1998,Takahashi1998}
and strong Coulomb interaction.
\cite{Imada1998a}
Because the charge incoherence conspicuously appears near the MI transition,
it is important to fully understand the undoped insulating states
and to consider doping effects on them in a unified framework
for treating quantum fluctuations.

In the present paper, ground state properties
of the extended DE model containing the Coulomb interaction,
twofold degeneracy of $e_{g}$ orbitals, and the JT coupling
are discussed.
The undoped and doped states
as well as critical properties of the MI transition
are examined in a unified framework.
A part of the results has been published
in the previous publication.
\cite{Motome1999a,Motome1999b}
In this paper, more systematic analyses are presented
including the JT effects in hole-doped regions.
As mentioned above, the real materials show strong 2D anisotropy
at and around the undoped state,
which suggests the importance of quantum fluctuations.
The JT coupling to some extent suppresses these quantum fluctuations
and plays a crucial role to reproduce experiments.
In this work, the quantum fluctuation effects are investigated
in the 2D anisotropic structure beyond the mean-field results;
we use here an unbiased approach
by employing the quantum Monte Carlo method.
\cite{White1989,Imada1989,Sorella1991}
Effects of the JT distortion are also studied by optimizing
the ground state of the Hamiltonian
with an electron-phonon coupling.
Physical properties in the optimized ground state are calculated
and compared with the experimental results
with emphasis on roles of the hierarchy structure in energy scales.
We determine values of model parameters
appropriate for the Mn compounds
and compare with those in other theoretical calculations.
For doping of carriers, 
critical enhancement of both orbital correlation length
and charge compressibility is obtained
in the absence of the JT distortion.
The results are examined to understand
the puzzling feature of the incoherent metal in experiments.
We also show that
the JT distortion rapidly becomes unstable with increasing doping.
Phase separation takes place in our model 
when we take account of the JT coupling in the doped cases,
which implies an instability to charge ordering in real materials.

This paper is organized as follows:
In Sec. \ref{Sec:Hamiltonian}, we describe derivation of our model 
and discuss the parameters used in the following discussions.
Numerical results are presented in Sec. \ref{Sec:Results}.
The numerical method and conditions are also described.
All the results are discussed in comparison with experiments
as well as with other theoretical studies in Sec. \ref{Sec:Discussion}.
Section \ref{Sec:Summary} is devoted to summary.

\section{Model}
\label{Sec:Hamiltonian}

In order to examine the above-mentioned experimental results
beyond the `simple' DE model defined in Eq. (\ref{H DE}),
we take account of on-site Coulomb interaction,
twofold degeneracy of conduction bands, and JT coupling.
Our Hamiltonian consists of three terms:
\cite{Motome1999a}
\begin{equation}
\label{H tot}
{\cal H} = {\cal H}_{\rm el} + {\cal H}_{\rm el-ph} + {\cal H}_{\rm ph}.
\end{equation}
The first electronic term ${\cal H}_{\rm el}$
is derived from the extended DE model
which contains the Coulomb interaction 
and two orbital degrees of freedom of $e_{g}$ electrons 
as described in Sec. \ref{Sec:H el} in detail.
The second and third terms denote the electron-phonon interaction
(JT coupling) and the elastic energy of phonons, respectively.
Phonons are treated in a classical way
as mentioned in Sec. \ref{Sec:el-ph coupling}.
Section \ref{Sec:Parameters} describes details of
the model parameters used in the following calculations.
In Sec. \ref{Sec:what parameter},
the derivation of our model is examined
to clarify energy scales and doping regions
where our model can be applied.

\subsection{Electronic part}
\label{Sec:H el}

The extended DE model containing
twofold degeneracy of $e_{g}$ orbitals 
as well as on-site Coulomb interaction is given by
\begin{equation}
\label{H DE2}
{\cal H}_{\rm DE}' = {\cal H}_{\rm hop} + {\cal H}_{\rm int}
+ {\cal H}_{K} + {\cal H}_{\mu},
\end{equation}
where
$
{\cal H}_{\rm hop} = \sum_{ij, \nu\nu', \sigma} 
t_{ij}^{\nu\nu'} c_{i\nu\sigma}^{\dagger} c_{j\nu'\sigma}
$
which describes electron hoppings;
${\cal H}_{\rm int}$ denotes on-site electronic interactions
within doubly-degenerate $e_{g}$ orbitals;
${\cal H}_{K}$ is similar to the second term in Eq. (\ref{H DE}) which
describes the Hund's rule coupling between
electrons in the $e_{g}$ orbitals and 
localized spins in the $t_{2g}$ orbitals; and
${\cal H}_{\mu}$ controls total electron density
by the chemical potential $\mu$.
The indices $i,j = 1,\cdot\cdot\cdot,N_{\rm S}$ and 
$\nu,\nu' = 1,2$ denote the lattice sites of Mn atoms and
the orbitals, respectively.
We take $\nu=1 (2)$ for the $d_{x^{2}-y^{2}} (d_{3z^{2}-r^{2}})$ orbital.
Although the $t_{2g}$-spin interaction 
such as $\mbox{\boldmath $S$}_{i} \cdot \mbox{\boldmath $S$}_{j}$
has been considered in several previous studies,
\cite{Ishihara1997a,Feiner1999,Maezono1998}
it is ignored in Eq. (\ref{H DE2})
because the energy scale is small.
This point will be discussed in Sec. \ref{Sec:what parameter}.

Following several previous studies,
\cite{Anderson1955,deGennes1960}
we consider the limit of the strong Hund's-rule coupling,
$K |\mbox{\boldmath $S$}^{t_{2g}}| \gg t_{ij}^{\nu\nu'}$,
which may be appropriate for situations in many perovskite manganites.
Since $\mbox{\boldmath $\sigma$}_{i}$ aligns parallel to 
$\mbox{\boldmath $S$}_{i}$ in each site in this limit,
the hopping integral between sites $i$ and $j$
is renormalized by a relative angle of the spins at sites $i$ and $j$.
Then the model (\ref{H DE2}) is simplified into
a spinless fermion model as
\begin{eqnarray}
{\cal H}_{\rm DE}' &\simeq&
\sum_{ij, \nu\nu'} \lambda_{ij} t_{ij}^{\nu\nu'}
c_{i\nu}^{\dagger} c_{j\nu'} \nonumber \\
&+&
\tilde{U} \sum_{i} (n_{i1}-\frac{1}{2}) (n_{i2}-\frac{1}{2})
+ \mu \sum_{i, \nu} n_{i\nu},
\label{H DE2 limit}
\end{eqnarray}
where
$\lambda_{ij}$ is a renormalization factor determined by
the expectation value of $\mbox{\boldmath $S$}_{i}$ and $\mbox{\boldmath $S$}_{j}$;
and $n_{i\nu} = c_{i\nu}^{\dagger} c_{i\nu}$ which denotes
the number operator.
Note that $\tilde{U}$ is not a bare on-site Coulomb interaction
but a renormalized one; $\tilde{U} = U_{12}-J_{12}$ where
$U_{12}$ is the interorbital Coulomb interaction
and $J_{12}$ is the Hund's rule coupling
between the orbitals $\nu=1$ and $2$.

In undoped state where one electron occupies a site
on average in the model (\ref{H DE2 limit})
(we call this situation half filling hereafter),
perovskite manganites commonly show
the $A$-type antiferromagnetism.
This state has strong 2D anisotropy attributed to
antiferromagnetic stacking of ferromagnetic planes ($xy$ planes).
\cite{Wollan1955,Goodenough1955,Hirota1996}
When we dope holes,
the system gradually recovers the three dimensionality
and finally becomes an isotropic ferromagnetic metal
through structural, magnetic and MI transitions.
\cite{Jonker1950,Santen1950,Wollan1955,Goodenough1955,Matsumoto1970b,Tokura1994,Urushibara1995,Kawano1996a,Kawano1996b}
However, the perfect polarization of spins in the 2D planes
persists through these transitions;
moreover, the 2D anisotropy remains for finite doping
at least until the magnetic or the structural transition.
\cite{Wollan1955,Goodenough1955,Matsumoto1970b,Tokura1994,Urushibara1995,Kawano1996a,Kawano1996b,Yamada1996,Zhou1997}
In this situation, the spin polarized model is well justified
within the 2D plane without the aposteriori and ambiguous
factor $\lambda_{ij}$.

On the basis of these considerations,
here we consider the spin-polarized 2D model
as a starting point to discuss physical properties
of manganites at and around half filling.
The explicit form of it is given by
\begin{eqnarray}
{\cal H}_{\rm el} &=&
\sum_{ij, \nu\nu'} \tilde{t}_{ij}^{\nu\nu'}
c_{i\nu}^{\dagger} c_{j\nu'} \nonumber \\
&+&
\tilde{U} \sum_{i} (n_{i1}-\frac{1}{2}) (n_{i2}-\frac{1}{2})
+ \mu \sum_{i, \nu} n_{i\nu}.
\label{H el}
\end{eqnarray}
Here we consider only the nearest-neighbor hopping integrals
which are explicitly given by
$\tilde{t}_{ij}^{11}=-3\tilde{t}_{0}/4$, $\tilde{t}_{ij}^{22}=-\tilde{t}_{0}/4$,
$\tilde{t}_{ij}^{12}=\tilde{t}_{ij}^{21}=-(+)\sqrt{3}\tilde{t}_{0}/4$
in the $x$($y$) direction.
\cite{Anderson1959}
The unit of hopping integrals $\tilde{t}_{0}$
is defined in Sec. \ref{Sec:el-ph coupling}.
Our assumptions to derive Eq. (\ref{H el}) are examined
in Sec. \ref{Sec:what parameter} in detail.

\subsection{Jahn-Teller coupling}
\label{Sec:el-ph coupling}

The Mn$^{3+}$ ion has an instability of a distortion
of MnO$_{6}$ octahedron.
This requires a strong electron-phonon coupling in the model,
which is the so-called JT coupling.
\cite{Kanamori1960}
There are two phonon modes which strongly couple with 
$e_{g}$ orbital degrees of freedom.
\cite{VanVleck1939,Goodenough1963}
In this work, lattice degrees of freedom are treated
in a classical or addiabatic approximation, that is,
kinetic energy of phonons is neglected.
\cite{Millis1996b,Millis1996a}
Then, the Hamiltonian for this JT coupling is given in the form
\begin{equation}
\label{H el-ph}
{\cal H}_{\rm el-ph} = -g \sum_{i} \mbox{\boldmath $u$}_{i} \cdot \mbox{\boldmath $I$}_{i},
\end{equation}
where $g$ is the electron-phonon coupling constant;
$\mbox{\boldmath $u$}$ and $\mbox{\boldmath $I$}$ are two-component vectors defined as
\begin{eqnarray}
\mbox{\boldmath $u$}_{i} &=& u_{i} \left( \cos 2\theta_{i}, \sin 2\theta_{i} \right) \\
\mbox{\boldmath $I$}_{i} &=& 2 \left( T_{i}^{z}, T_{i}^{x} \right),
\end{eqnarray}
where $u_{i}$ denotes a displacement of the oxygens
surrounding $i$-th Mn site and
$\theta_{i}$ is a coupling angle
defined later in Eq. (\ref{gamma def}).
$T_{i}^{\mu} (\mu=x,y,z)$ is a pseudo-spin operator which denotes
two degrees of freedom of the degenerate orbitals as
\begin{equation}
\label{T def}
T_{i}^{\mu} = \frac{1}{2} \sum_{\nu\nu'} \mbox{\boldmath $\tau$}_{\nu\nu'}^{\mu}
c_{i\nu}^{\dagger} c_{i\nu'},
\end{equation}
where $\mbox{\boldmath $\tau$}$ is the Pauli matrix.

The term given by Eq. (\ref{H el-ph}) is rewritten 
by substituting the definition (\ref{T def}) as
\begin{equation}
        \label{H el-ph2}
{\cal H}_{\rm el-ph} =
-g \sum_{i} u_{i} \left( m_{i1} - m_{i2} \right),
\end{equation}
where $m_{i\nu} \equiv \gamma_{i\nu}^{\dagger} \gamma_{i\nu}$
is the number operator for the new operator $\gamma$
which is a linear combination of the operator $c$ defined as
\begin{equation}
\label{gamma def}
\left( \begin{array}{c}
       \gamma_{i1} \\ \gamma_{i2}
       \end{array} \right)
=
\left( \begin{array}{cc}
          \cos \theta_{i} & \sin \theta_{i} \\
          -\sin \theta_{i} & \cos \theta_{i}
          \end{array} \right)
\left( \begin{array}{c}
          c_{i1} \\ c_{i2}
          \end{array} \right).
\end{equation}
Then, Eq. (\ref{H el-ph2}) shows that the JT coupling
in the classical approximation can be expressed as the local fields
which lift the degeneracy of two orbitals 
diagonalized in the rotated basis
by the coupling angle $\theta_{i}$ as in Eq. (\ref{gamma def}).

The last term in the Hamiltonian (\ref{H tot}) denotes
the elastic energy of phonons as
\begin{equation}
        {\cal H}_{\rm ph} = k \sum_{i} u_{i}^{2},
\end{equation}
where $k$ is the spring constant.

When an oxygen connecting $i$ and $j$-th Mn sites is displaced
by an amplitude $u_{i}$, the hopping integral
between Mn sites, $\tilde{t}_{ij}^{\nu\nu'}$, is modulated
since it originally consists of a product of
hopping integrals between Mn and O sites.
Note that in both the phonon modes considered here,
oxygens shift only in the direction of Mn-O-Mn bonds.
The Mn-O hopping is modified as $\left(1 \pm 2u_{i}\right)^{-7/2}$
under the displacement of $u_{i}$
\cite{Harrison1980}
in the length unit
mentioned in the next subsection Sec. \ref{Sec:Parameters}.
Therefore,
we take the unit of Mn-Mn hopping integrals in Eq. (\ref{H el}) as
\begin{equation}
\tilde{t}_{0} = t_{0}
\left( 1 - 4 u_{i}^{2} \right)^{-7/2}.
\end{equation}

\subsection{Model parameters}
\label{Sec:Parameters}

We set $t_{0}$ and the Mn-Mn lattice constant
as the energy unit and the length unit, respectively.
Our Hamiltonian contains the following parameters;
the Coulomb interaction $\tilde{U}$,
the electron-phonon coupling constant $g$,
the spring constant $k$,
and the parameters $u_{i}$ and $\theta_{i}$ which determine
local JT distortions.
From experiments and band calculations,
\cite{Bocquet1992,Chainani1993,Arima1993,Saitoh1995,Satpathy1996,Solovyev1996a,Solovyev1996b}
$t_{0}$ is considered to be around $0.5$eV and
$\tilde{U}$ is estimated as several eV.
In this work, $\tilde{U}/t_{0}$ is considered as a parameter
in the following calculations.
From our numerical results in Sec. \ref{Sec:Results},
the realistic value of $\tilde{U}$ is discussed
and compared with that in other theoretical investigations
in Sec. \ref{Sec:Discussion}.
The spring constant $k$ is roughly estimated from
the frequency of an oxygen bond stretching phonon as
the order of $10 \sim 100$ eV; we use here $k=100$.
\cite{Okimoto1995,Okimoto1997,Millis1996b}
The electron-phonon coupling constant $g$
is hard to estimate from available experimental data.
It may, however, be the order of $1$eV;
we take $g=10$ in this work.
The values of $k$ and $g$ are examined in Sec. \ref{Sec:Mott}
by quantitative comparison between our results and experiments.
The parameters $u_{i}$ and $\theta_{i}$ are determined
to minimize the total energy.
For simplicity, they are treated
in a mean-field scheme as described in Sec. \ref{Sec:HFwithJT} in detail.

\subsection{Energy scale and doping range}
\label{Sec:what parameter}

In the following, the model given by Eq. (\ref{H tot})
is considered in two dimensions.
Here, we discuss the energy scale and
the region of the doping concentration $\delta$
where the present model is justified.
In particular, two main simplifications
in the model (\ref{H el}) are examined;
one is neglect of three dimensionality and
the other is the full polarization of spins.
We also comment on the $t_{2g}$-spin exchange interaction
neglected in Eq. (\ref{H DE2}).

In undoped states, the largest energy scale in our consideration
is the charge gap $\Delta_{\rm c}$ of the order of $1$eV.
The formation of the gap is a common feature
irrespective of the dimensionality.
Therefore, physics of the charge gap formation
in the energy scale of $\Delta_{\rm c}$ may well be captured
in the 2D model equally to the isotropic 3D system.

When the energy scale is lowered below $1$eV,
the system becomes insulating and excitations are described
through spin and orbital exchange processes
because single-particle charge fluctuations are suppressed.
Through these processes, the JT distortion and the orbital ordering
take place in the energy scale of the order of $0.1$eV.
Although the interlayer structure of JT distortion and orbital ordering
cannot be predicted, the energy scale of the distortion and
the orbital ordering are again captured by the 2D model,
because they take place in a planar configuration and
the energy scale of these orderings may primarily be estimated
from such anisotropic structure.

Although we employ a spin-polarized model,
the spin configuration is not an important issue in this energy scale
because the spin exchange energy is too small
as compared to the above-mentioned energy scales.
Whether spins are ordered or not does not make
any crucial difference in our discussions.

With further lowering of the energy scale to the order of $0.01$eV,
a shortcoming of the spin-polarized 2D model becomes clear
because the 3D spin ordering takes place
under a competition of different possible 3D structures.
We of course are not able to discuss such process.
However, eventually, the $A$-type antiferromagnetic structure
under the 2D anisotropic orbital ordering takes place,
which again makes spin-polarized 2D model useful
due to the strongly anisotropic structure
with ferromagnetic alignment of spins in each plane
at very low temperatures $\sim 10$K.
We emphasize that the above energy hierarchy in undoped states
provides the underlying backbone structure of this problem.
Our model is useful in all the energy scales
except those of the 3D spin-ordering process.

We neglect the $t_{2g}$-spin interaction in the starting DE model
given by Eq. (\ref{H DE2}).
The exchange coupling between nearest-neighbor $t_{2g}$ spins
is roughly estimated at the order of $0.001$ or $0.01$eV.
\cite{Goodenough1955}
This energy scale is in the same order as 
the 3D magnetic transition temperature $T_{\rm N}$.
Actually, the $t_{2g}$-spin interaction is considered to be
one of the origins of the $A$-type antiferromagnetism
in undoped states.
\cite{Ishihara1997a,Feiner1999,Maezono1998}
Following the above discussions on the energy scales,
we consider that
the inclusion of this interaction term will not affect
the essence of the following results in this paper.

Upon doping, 
a canted antiferromagnetic state is experimentally observed
near the undoped state.
\cite{Wollan1955,Matsumoto1970b,Kawano1996a,Kawano1996b}
In this state, the perfect polarization of spins in each plane
remains unperturbed, however, the direction of polarization
tilts to the $z$ direction.
The canting angle increases with the increase in the doping concentration $\delta$ and
finally the isotropic ferromagnetic state appears with MI transition.
This recovers the three dimensionality.
When we discuss the order-disorder transitions
of the JT distortion and the orbital ordering,
this 3D effect may be important.

We here estimate energy scales of the 3D effect by the spin canting.
One is given by the transition temperature $T_{\rm N}$.
In La$_{1-x}$Sr$_{x}$MnO$_{3}$, $T_{\rm N}$ does not strongly
depend on the doping concentration and
remains the order of $0.01$eV through the spin-canted phase.

Another energy scale is kinetic energy in the $z$ direction
gained with the increase in the canting angle.
We here estimate this energy gain following the work by de Gennes.
\cite{deGennes1960}
He analyzed the DE model (\ref{H DE}) by mean-field approximation
to explain the origin of spin-canted phase
observed by neutron scattering experiment
for La$_{1-x}$Ca$_{x}$MnO$_{3}$.
If his formalism is applied straightforwardly,
roughly speaking, the kinetic-energy gain of each hole
increases linearly with the canting angle and hence
linearly with the doping concentration.
Then, the total kinetic-energy gain is estimated as
\begin{equation}
\label{deltaEtz}
\Delta E_{t}^{z} \simeq -\frac{(\tilde{t}^{22})^{z}}{\delta_{\rm m}}
\delta^{2},
\end{equation}
for $0 < \delta \simle \delta_{\rm m}$
where $\delta_{\rm m}$ is the critical concentration for
the ferromagnetic transition.
In the case of La$_{1-x}$Sr$_{x}$MnO$_{3}$,
the transition occurs at $\delta_{\rm m} \simeq 0.16$.
\cite{Kawano1996a,Kawano1996b}
In terms of our model, $(\tilde{t}^{22})^{z}$ is equal to $-\tilde{t}_{0}$.

In order to discuss the 2D anisotropic structure of
the JT distortion and the orbital ordering
in the energy scales smaller than $T_{\rm N}$ or $\Delta E_{t}^{z}$
in doped states,
the three dimensionality should be taken into account.
In other words, our 2D model may be useful to discuss 
these ordering processes in the energy scales
larger than $T_{\rm N}$ or $\Delta E_{t}^{z}$
even at finite hole doping.

In the metallic state with further doping, 
if we assume that the MI transition is
a continuous or a weak first-order one,
there is a critical region of $\delta$
where the system is controlled by the criticality of the transition.
In the real materials, the insulators show the 2D anisotropy
in both the orbital ordering and the JT distortion.
The metallic state near the MI transition
may be influenced by the 2D anisotropy of the insulator.
\cite{Imada1998a}
Our 2D model would serve
to discuss the criticality of the MI transition.

\section{Results}
\label{Sec:Results}

This section presents numerical results for the model (\ref{H tot})
in the ground state.
First, Sec. \ref{Sec:Method} describes numerical technique and
details of computational conditions.
Then in Sec. \ref{Sec:Half filling},
we show numerical results at half filling without and with the JT coupling.
Results for hole-doped cases are presented in Sec. \ref{Sec:Hole doping}.
All the results are discussed
in comparison with experiments as well as with other theoretical
investigations in Sec. \ref{Sec:Discussion}.

\subsection{Method and computational details}
\label{Sec:Method}

We investigate ground state properties of the model (\ref{H tot})
by the projection quantum Monte Carlo (PQMC) method.
\cite{White1989,Imada1989,Sorella1991}
If JT distortions are given, that is,
if the parameters $u_{i}$ and $\theta_{i}$ are fixed,
the Hamiltonian given by Eq. (\ref{H tot}) is the Hubbard model
with a generalized hopping and external fields.
Therefore we can apply a well-established algorithm for
the ordinary single-band Hubbard model straightforwardly.
\cite{Hirsch1983,Hirsch1985,White1989,Imada1989,Sorella1991}
In two dimensions, quantum fluctuations may be important
even in insulating states.
PQMC gives unbiased results on the effects of electron correlation
and quantum fluctuations.

For simplicity,
the JT coupling given by Eq. (\ref{H el-ph})
is treated in a mean-field scheme.
As described in Sec. \ref{Sec:HFwithJT},
we fix $\theta_{i}$ following the experimental indications
and find a uniform solution $u_{i}=u$ to minimize total energy by PQMC.

For the present model, PQMC has the so-called negative sign problem
even at half filling with $g=0$,
although this problem is completely absent for
the ordinary Hubbard model at half filling.
\cite{Hirsch1983,Hirsch1985}
However, from practical point of view,
we have managed to reach the ground-state convergence
in several choices of parameter values.
Physical quantities presented in the following
are obtained in the parameter region
where the negative sign problem is not serious in
reaching the ground state properties with needed accuracy.
The details are discussed in Appendix A.

In PQMC, ground states are projected out
by operating $\exp(-\beta {\cal H})$ on a trial wave function.
For good convergence about the projection $\beta$,
we employ several different trial wave functions depending on the cases;
noninteracting Slater determinant,
unrestricted Hartree-Fock solutions,
\cite{Furukawa1991a}
and orbital-singlet wave functions
stabilized by small dimerization in hoppings.
\cite{Assaad1997}
The convergence for these trial wave functions is discussed
in Appendix A.
Discreteness $\Delta\tau$ of the slices in the projection $\beta$
is taken as $\Delta\tau t_{0} = 0.05$
which is small enough to obtain ground state properties within numerical errorbars.

The 2D systems with $N_{\rm S}=L^{2}$ sites are investigated in the following.
As shown in Appendix B, we note that this model defined on a finite-size lattice shows
different behavior depending on the linear dimension $L=4n$ or $4n+2$
($n$ is an integer) as well as on boundary conditions
due to the shapes of Fermi surfaces in the finite-size systems.
In extrapolations to the thermodynamic limit $N_{\rm S}\rightarrow \infty$,
we pay attention to choose a suitable set of $L$ and the boundary condition
to reduce finite-size effects.
In the following, we mainly show the results
for a series of $L=4n$ with the antiperiodic boundary condition
and $L=4n+2$ with the periodic boundary condition.
See Appendix B for details.

\subsection{Half filling}
\label{Sec:Half filling}

\subsubsection{Without Jahn-Teller coupling}
\label{Sec:HFwithoutJT}

First, the ground states at half filling are investigated
in the absence of the JT coupling.
The system becomes Mott insulating
for finite values of the Coulomb interaction $\tilde{U}$.
In the following, we show numerical results on
the charge gap amplitude and the orbital ordering at this filling.

\begin{figure}
\epsfxsize=8cm
\centerline{\epsfbox{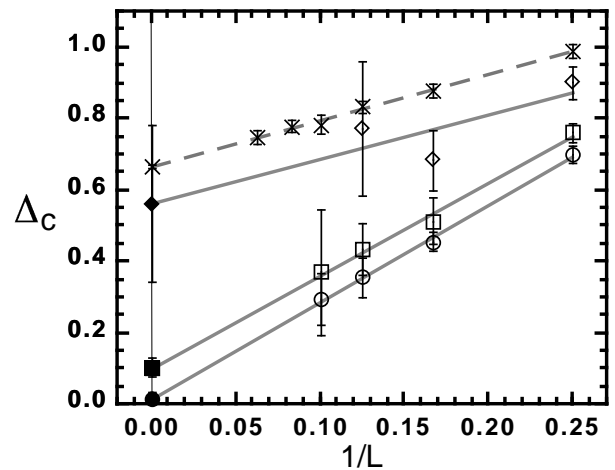}}
\caption{
System-size dependence of the charge gap amplitude at half filling.
The circles, squares, and diamonds correspond to 
$\tilde{U} =3, 4$, and $5$, respectively.
The data are fitted by $1/L$ and the extrapolated values
to $L \rightarrow \infty$ are plotted as the filled symbols.
The crosses show the data on the ordinary Hubbard model at $U = 4$.
\protect{\cite{Assaad1996a}}
}
\label{Fig:GapExtwithoutJT}
\end{figure}

Figure \ref{Fig:GapExtwithoutJT} shows the system-size dependence of
the charge gap amplitude for various values of
the Coulomb interaction $\tilde{U}$.
Here, the charge gap is calculated from
the shift of the chemical potential when we dope holes to the
half-filled state.
\cite{Furukawa1992}
The data for each value of $\tilde{U}$ are well fitted by $1/L$
as in the ordinary single-band Hubbard model.
\cite{Furukawa1992,Assaad1996a}

\begin{figure}
\epsfxsize=8cm
\centerline{\epsfbox{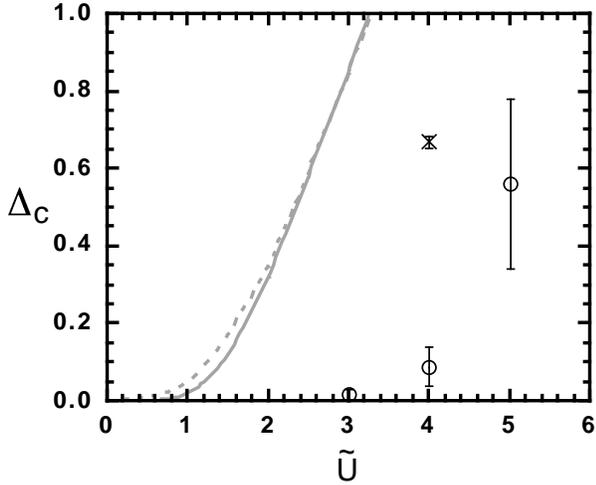}}
\caption{
$\tilde{U}$ dependence of the charge gap amplitude in the thermodynamic limit.
The circles and cross show the QMC data on the present model
and on the ordinary Hubbard model at $\tilde{U} = U$,
\protect{\cite{Assaad1996a}}
respectively.
The gray and dotted curves are the mean-field results
for the present model and the ordinary Hubbard model, respectively.
}
\label{Fig:GapwithoutJT}
\end{figure}

Figure \ref{Fig:GapwithoutJT} summarizes the charge gap amplitude
in the thermodynamic limit obtained by the extrapolations
in Fig. \ref{Fig:GapExtwithoutJT}.
For comparison, we plot the QMC results for the ordinary Hubbard model
\cite{Assaad1996a}
as well as the results of the mean-field approximations for both models.
The QMC data on the present model show remarkable reduction of the charge 
gap amplitude in the region of $\tilde{U}$ considered here
in comparison with the mean-field results as well as
the QMC result in the ordinary Hubbard model.
This means that quantum fluctuations may be more important in the present model
than in the ordinary Hubbard model.
Although further studies are necessary to discuss
the value of $\tilde{U}$ for the opening of the gap,
increasing the Coulomb interaction $\tilde{U}$ opens the charge gap and
let the system be Mott insulating; however in order to
reproduce the realistic values of charge gap amplitude
($\Delta_{\rm c} \sim 2t_{0}$)
only with $\tilde{U}$ without the JT coupling,
it is necessary to take much larger interaction
than $\tilde{U} = 5$.

\begin{figure}
\epsfxsize=8cm
\centerline{\epsfbox{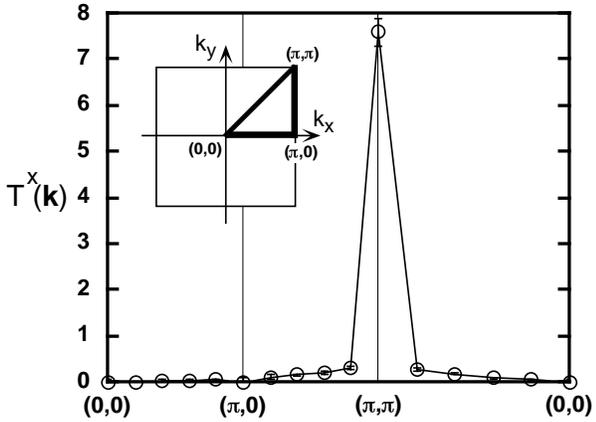}}
\caption{
The orbital correlation function ($x$ component)
at $\tilde{U} = 4$ in the system with $N_{\rm S} = 10\times 10$ at half filling.
The data are presented along the bold line in the $k$ space
as shown in the inset.
}
\label{Fig:Tx(k)}
\end{figure}

The Mott insulating state induced by
the Coulomb interaction shows a staggered orbital ordering.
We calculate the orbital correlation function defined as
\begin{equation}
        T^{\mu}\left(\mbox{\boldmath $k$}\right) = N_{\rm S}^{-1} \sum_{ij}
        T_{i}^{\mu} T_{j}^{\mu} \exp\left( {\rm i}\mbox{\boldmath $k$} 
        \cdot \mbox{\boldmath $r$}_{ij}\right)
        \label{Tkdef}
\end{equation}
with the operator given by Eq. (\ref{T def}) ($\mu=x,y,z$).
For a finite value of $\tilde{U}$,
$T^{x}(\mbox{\boldmath $k$})$ shows a peak at
$\mbox{\boldmath $k$}=\mbox{\boldmath $Q$}\equiv(\pi,\pi)$
as shown in Fig. \ref{Fig:Tx(k)},
while $T^{y}(\mbox{\boldmath $k$})$ and $T^{z}(\mbox{\boldmath $k$})$
do not have a conspicuous structure.

\begin{figure}
\epsfxsize=8cm
\centerline{\epsfbox{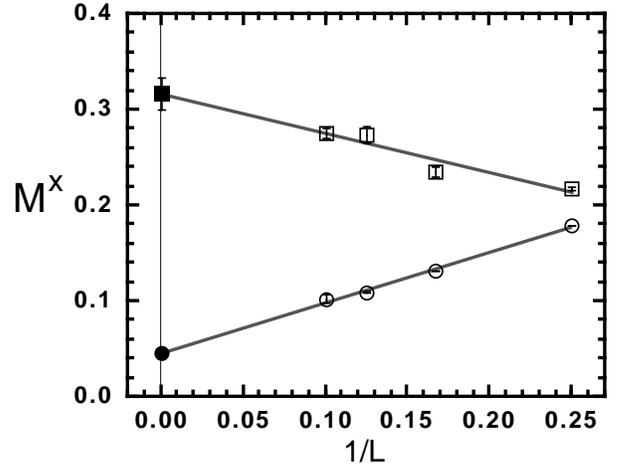}}
\caption{
System-size dependence of the moment of the orbital polarization
at half filling.
The circles and squares are for $\tilde{U} = 3$ and $4$, respectively.
The data are fitted by $1/L$ and the extrapolated values
to $L \rightarrow \infty$ are plotted as the filled symbols.
}
\label{Fig:MxExtwithoutJT}
\end{figure}

\begin{figure}
\epsfxsize=8cm
\centerline{\epsfbox{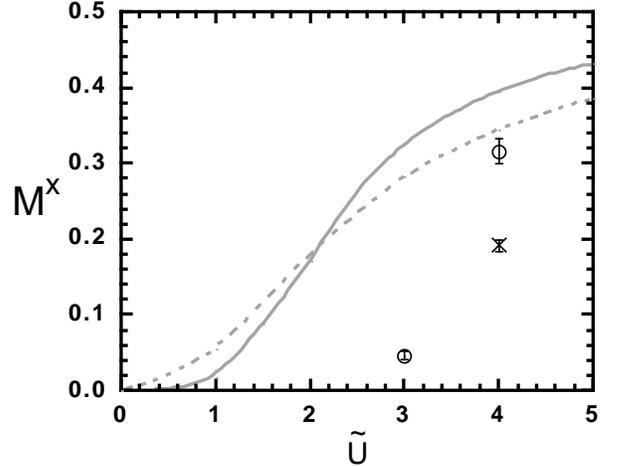}}
\caption{
$\tilde{U}$ dependence of the long-ranged order of
the orbital polarization in the thermodynamic limit.
The circles and cross show the QMC data on the present model
and for the ordinary Hubbard model at $\tilde{U} = U$,
\protect{\cite{White1989}}
respectively.
The gray and dotted curves are the mean-field results
for the present model and
for the long-ranged antiferromagnetic order of
the ordinary Hubbard model, respectively.
}
\label{Fig:MxwithoutJT}
\end{figure}

The moment of this staggered orbital ordering is calculated
by $M^{x} = [T^{x}(\mbox{\boldmath $Q$})/N_{\rm S}]^{1/2}$.
Figure \ref{Fig:MxExtwithoutJT} plots the system-size dependence of
the staggered moment $M^{x}$.
The data are well fitted by $1/L$, from which
we obtain the results in the thermodynamic limit
summarized in Fig. \ref{Fig:MxwithoutJT}.
We also plot in the figure the QMC result
for the antiferromagnetic spin-ordering moment
in the ordinary Hubbard model
\cite{White1989}
and the mean-field results as in Fig. \ref{Fig:GapwithoutJT}.
In small $\tilde{U}$ region,
the moment grows slower than the mean-field results
as the charge gap amplitude in Fig. \ref{Fig:GapwithoutJT},
however it grows rapidly around $\tilde{U}=4$;
this may be related to the uniaxial property of the moment in the present model.
Actually, in numerical calculations for small clusters,
\cite{NakanoPC}
the moment in our model grows rapidly and is saturated to its full value $M^{x}=1/2$
for large values of $\tilde{U}$;
on the other hand, the isotropic moment in the ordinary Hubbard model
slowly approaches a reduced value
because of stronger quantum fluctuations.

For the following discussions,
we consider here the pattern of the orbital ordering.
Our model exhibits a staggered ordering of
$( |d_{x^{2}-y^{2}}\rangle + |d_{3z^{2}-r^{2}}\rangle) /
( |d_{x^{2}-y^{2}}\rangle - |d_{3z^{2}-r^{2}}\rangle)$ type
indicated by the peak structure of $T^{x}(\mbox{\boldmath $Q$})$.
If we define a single-site state
by the linear combination of the two orbitals as
\begin{equation}
        \label{lc of orbital}
        |\psi (\phi_{i}) \rangle = |d_{x^{2}-y^{2}}\rangle \cos \phi_{i}
        + |d_{3z^{2}-r^{2}}\rangle \sin \phi_{i},
\end{equation}
the ordering pattern is characterized by
$\phi_{i} = (-1)^{|\mbox{\boldmath $r$}_{i}^{x} + 
\mbox{\boldmath $r$}_{i}^{y}|} \pi/4$.
See Fig. \ref{Fig:phi}.
The previous mean-field studies have predicted slightly different patterns
from our results, $\phi_{i} \simeq (-1)^{|\mbox{\boldmath $r$}_{i}^{x} + 
\mbox{\boldmath $r$}_{i}^{y}|} \pi/3$.
\cite{Kugel1973,Shiina1997,Feiner1999,Maezono1998,Mizokawa1995,Mizokawa1996}
We will discuss this discrepancy as a 3D effect
in Sec. \ref{Sec:Pattern}.

\subsubsection{With Jahn-Teller coupling}
\label{Sec:HFwithJT}

Next, effects of the JT coupling at half filling are investigated.
In experiments, the cooperative JT distortions are observed with 2D anisotropy
whose ordering pattern is characterized by
\begin{equation}
        \theta_{i} = (-1)^{|\mbox{\boldmath $r$}_{i}^{x} + \mbox{\boldmath $r$}_{i}^{y}|} \pi / 6
        \label{thetadef}
\end{equation}
with a uniform amplitude of $u \sim 0.035$
in terms of Eq. (\ref{H el-ph}).
\cite{Ellemans1971}
Following these results,
we treat here lattice degrees of freedom in a mean-field scheme for simplicity;
we fix $\theta_{i}$ as Eq. (\ref{thetadef}), and
optimize the total energy by PQMC to determine a uniform solution of $u$.

\begin{figure}
\epsfxsize=8cm
\centerline{\epsfbox{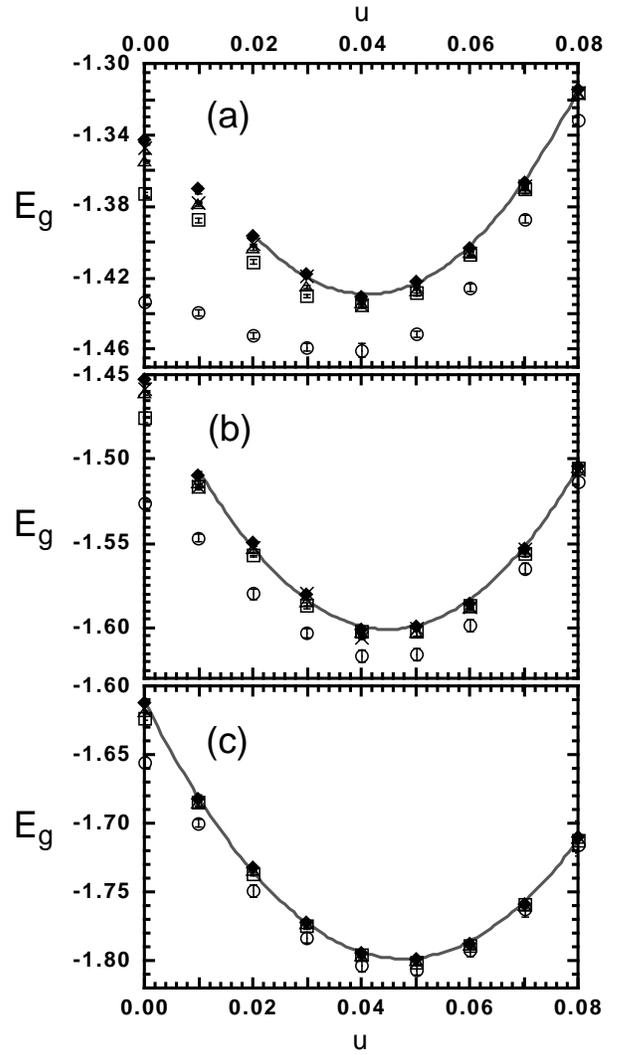}}
\caption{
$u$ dependence of the ground state energy at half filling;
(a) $\tilde{U} = 3$, (b) $\tilde{U} = 4$ and (c) $\tilde{U} = 5$.
We take $g=10$ and $k=100$.
The circles, squares, triangles, crosses, and diamonds correspond
to the data for $L = 4, 6, 8, 10$, and $\infty$, respectively.
The data for $L = \infty$ are obtained by the extrapolation
shown in Fig. \protect{\ref{Fig:EgExt}}.
The curves show the fits by Eq. ~(\protect{\ref{Egfit}})
for the data for $L = \infty$.
}
\label{Fig:Egvsu}
\end{figure}

\begin{figure}
\epsfxsize=8cm
\centerline{\epsfbox{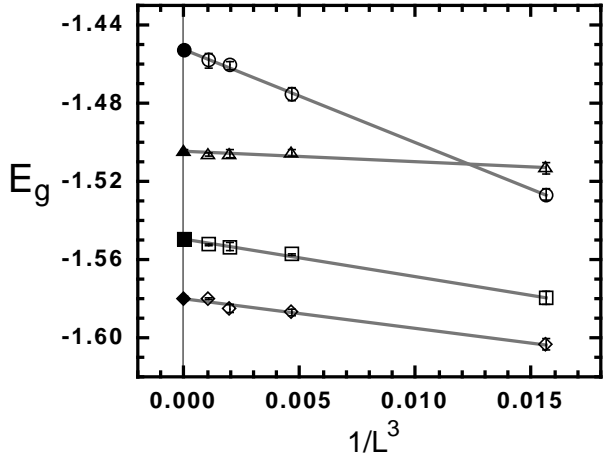}}
\caption{
System-size extrapolation of the ground state energy for $\tilde{U} = 4$
at half filling.
The circles, squares, diamonds, and triangles correspond to
the data for $u = 0.0, 0.02, 0.03$, and $0.08$, respectively.
The data are fitted by $1/L^{3}$ and the extrapolated values
to $L \rightarrow \infty$ are plotted as the filled symbols.
}
\label{Fig:EgExt}
\end{figure}

\begin{figure}
\epsfxsize=8cm
\centerline{\epsfbox{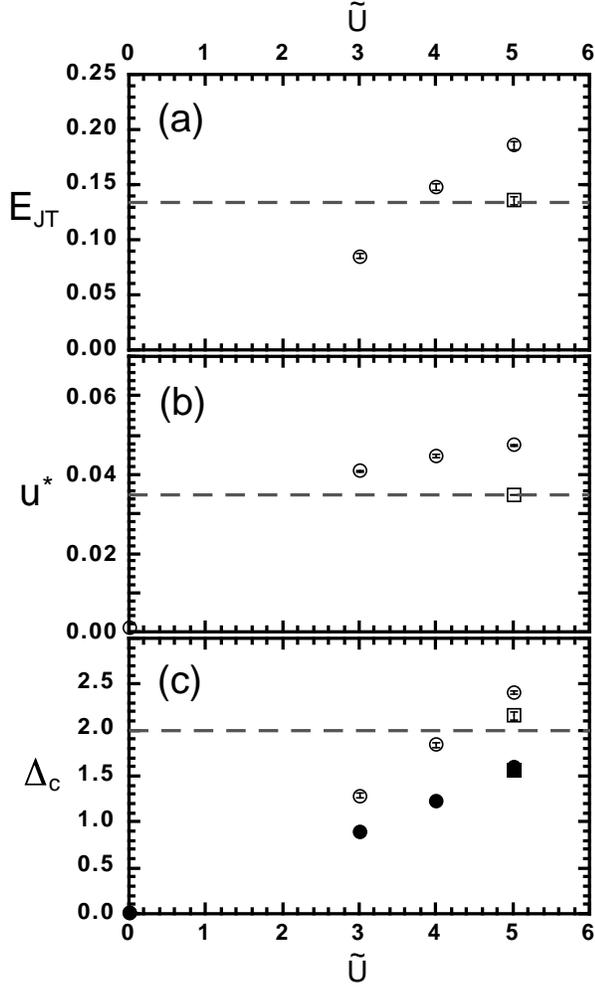}}
\caption{
$\tilde{U}$ dependence of (a) the stabilization energy of JT distortions,
(b) the amplitude of oxygen distortions, and
(c) the charge gap amplitude.
We take $g=10$ and $k=100$.
In (c), the open (filled) symbols denote the charge gap for
the electron (hole) doping.
The dotted lines show the experimental results of
the corresponding quantities.
The squares denotes the data at $g=10$ and $k=130$.
See the text for details.
}
\label{Fig:GapEJTuvsU}
\end{figure}

Figure \ref{Fig:Egvsu} illustrates these procedures.
We change the values of $u$ and calculate the total energy by PQMC
to find the optimized ground state.
In the figures, the data at $N_{\rm S} = \infty$ are obtained by
extrapolations by using the size scaling relation
$E_{\rm g} (L) \simeq E_{\rm g}(L=\infty) + a L^{-3}$
based on the spin-wave theory,
\cite{Huse1988,White1989}
as exemplified in Fig. \ref{Fig:EgExt}.
For each value of $\tilde{U}$, the total energy is considerably lowered
by introducing JT distortions and
the data are well fitted by
\begin{equation}
        \label{Egfit}
        E_{\rm g}(u) \simeq E_{\rm g}(u=0) - E_{\rm JT}
        + a (u-u^{*})^{2}
\end{equation}
near the minimum,
where $E_{\rm JT}$ is the stabilization energy of the JT distortion 
and
$u^{*}$ is the optimized amplitude of oxygen displacement.
The data on $E_{\rm JT}$ and $u^{*}$ are summarized
in Fig. \ref{Fig:GapEJTuvsU} and will be discussed below.

\begin{figure}
\epsfxsize=8cm
\centerline{\epsfbox{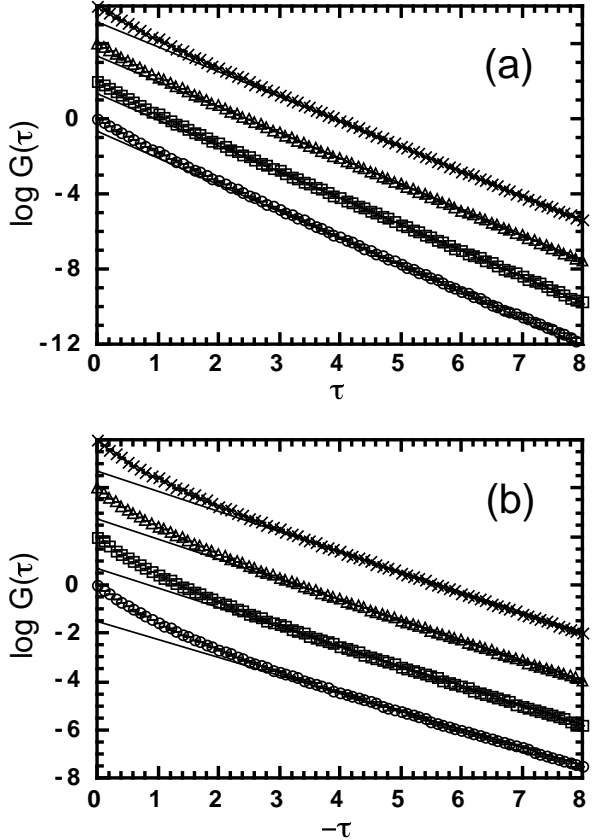}}
\caption{
Imaginary-time dependence of the uniform Green's function
for the ground state at $\tilde{U}=3$, $g=10$, and $k=100$:
(a) $\tau > 0$ and (b) $\tau < 0$.
The circles, squares, triangles, and crosses correspond to
the data for $L = 4, 6, 8$, and $10$, respectively.
Note that the origin for the $\log G(\tau)$ has
an offset to distinguish the data for each system size
in both figures.
The lines are the fits to the exponential tails
to calculate the charge gap amplitude.
}
\label{Fig:Gtau}
\end{figure}

\begin{figure}
\epsfxsize=8cm
\centerline{\epsfbox{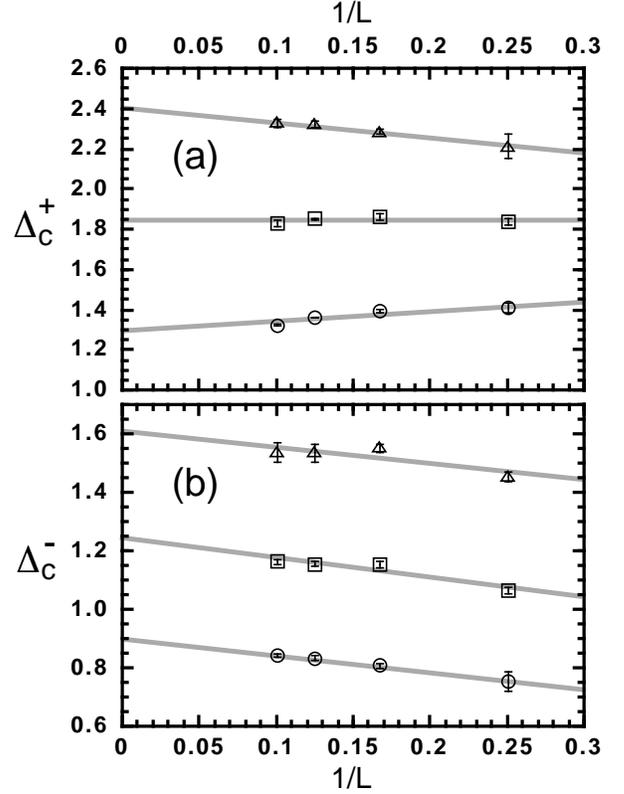}}
\caption{
System-size dependence of the charge gap amplitude for
(a) electron doping and (b) hole doping.
The circles, squares, and triangles correspond to
the data at $\tilde{U} = 3, 4$, and $5$, respectively.
The lines are the fits of $1/L$.
}
\label{Fig:GapExtwithJT}
\end{figure}

The JT distortions reduce quantum fluctuations and
stabilize the insulating states.
We calculate the charge gap amplitude for the optimized states
at $u=u^{*}$.
The JT coupling given by Eq. (\ref{H el-ph}) breaks the particle-hole symmetry at half filling,
therefore we calculate the charge gap for both
electron doping ($\Delta_{\rm c}^{+}$) and hole doping ($\Delta_{\rm c}^{-}$).
Here, the charge gap is calculated
from the imaginary-time dependence of
uniform single-particle Green's functions defined as
$G(\mbox{\boldmath $r$}_{ij}=0,\tau) \equiv \sum_{\nu} G_{\nu}(\mbox{\boldmath $r$}_{ij}=0,\tau)$ with
\begin{eqnarray}
        G_{\nu}(\mbox{\boldmath $r$}_{ij},\tau) &=& \Theta(\tau)
        \langle \Psi_{0} | c_{j\nu}(\tau) c_{i\nu}^{\dagger} | \Psi_{0} \rangle
         \nonumber \\
        &-& \Theta(-\tau)
        \langle \Psi_{0} | c_{i\nu}^{\dagger}(\tau) c_{j\nu} | \Psi_{0} \rangle,
\end{eqnarray}
where $| \Psi_{0} \rangle$ is the normalized ground state
at half filling and
$\Theta(\tau)$ is the step function.
The recently-developed stable algorithm is applied to obtain
$G(\tau)$ for the large values of the imaginary time $|\tau|$.
\cite{Assaad1996a}
The results of $G(\tau)$ are shown in Fig. \ref{Fig:Gtau}
for the case of $\tilde{U}=3$ as an example.
The exponential fits to the tails of $G(\tau>0)$ ($G(\tau<0)$)
give the values of
the charge gap $\Delta_{\rm c}^{+}$ ($\Delta_{\rm c}^{-}$).
Figure \ref{Fig:GapExtwithJT} shows the system-size dependence of
the charge gap amplitude obtained
from the fits in Fig. \ref{Fig:Gtau}.
The extrapolation in the $1/L$ dependence gives 
the data in the thermodynamic limit.
The amplitudes of the charge gap are remarkably enhanced
by the JT distortions from the values in Fig. \ref{Fig:GapwithoutJT}.

Figure \ref{Fig:GapEJTuvsU} summarizes our results on the half-filled states
in the presence of both the Coulomb interaction and the JT coupling.
We plot the stabilization energy of the JT distortion $E_{\rm JT}$,
the amplitude of the oxygen displacement $u^{*}$,
and the charge gap amplitude $\Delta_{\rm c}^{\pm}$
as a function of the Coulomb interaction $\tilde{U}$.
The corresponding values in experiments on LaMnO$_{3}$
are also shown for comparison:
For the stabilization energy $E_{\rm JT}$,
we show the ordering temperature of orbital and lattice $\simeq 780$K.
\cite{Wollan1955,Goodenough1955,Matsumoto1970a,Murakami1998}
The temperature is not exactly the same quantity as $E_{\rm JT}$, however
it is useful to compare these quantities at least in their orders of magnitude.
This point will be discussed later in Sec. \ref{Sec:Hierarchy}.
For the oxygen displacement $u^{*}$,
we cite X-ray experiment which gives $u^{*} \simeq 0.035$.
\cite{Ellemans1971}
For the charge gap amplitude, we show the experimental value
$\Delta_{\rm c} \simeq 1$eV obtained by the optical conductivity.
\cite{Arima1993}
As shown in Figs. \ref{Fig:GapEJTuvsU}(a)-(c),
these experimental results are consistently reproduced in our calculations
for the value of $\tilde{U} \simeq 4$ or $5$,
although slight adjustment on $\tilde{U}$, $g$ or $k$ is required
for more quantitative argument.
For instance, if we take $k=130$ at $\tilde{U}=5$ and $g=10$,
as shown in Fig. \ref{Fig:GapEJTuvsU},
we obtain more quantitative agreement
between the numerical data and the experimental values.
These results will be discussed
in Sec. \ref{Sec:Hierarchy} in more detail.

\begin{figure}
\epsfxsize=8cm
\centerline{\epsfbox{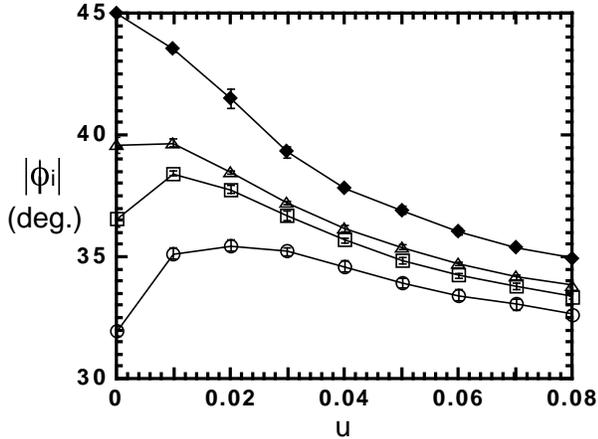}}
\caption{
$u$ dependence of the angle $\phi_{i}$
defined in Eq. ~(\protect{\ref{lc of orbital}}).
We take $\tilde{U}=4$, $g=10$, and $k=100$.
The circles, squares, triangles, and diamonds correspond to
the data for $L=4, 6, 8$, and $\infty$, respectively.
The lines connecting the data are guides to eye.
}
\label{Fig:anglephi}
\end{figure}

The JT coupling with the angle of Eq. (\ref{thetadef}) changes
the orbital-ordering pattern.
Figure \ref{Fig:anglephi} shows the angle of the pattern, $\phi_{i}$
defined in Eq. (\ref{lc of orbital}), as a function of $u$.
The angle $|\phi_{i}|$ is calculated from the moments of
orbital polarizations, $M^{x}$ and $M^{z}$,
in the form
\begin{equation}
|\phi_{i}| = \frac{1}{2} \arctan \left( \frac{M^{x}}{M^{z}} \right),
\end{equation}
which is easily confirmed by the definition (\ref{lc of orbital}).
Note that the case of $u=0$ corresponds to the result
in Sec. \ref{Sec:HFwithoutJT}, that is, $|\phi_{i}| = \pi/4$.
For the large values of $u$, the angle $|\phi_{i}|$ approaches
$\pi/6$ which equals to $|\theta_{i}|$ in Eq. (\ref{thetadef}).
In the case of $\tilde{U} = 4$, $g=10$, and $k=100$,
the angle $|\phi_{i}|$ is around $\pi/5$ for $u^{*} \simeq 0.0475$.
See also Fig. \ref{Fig:phi}.
This result will be compared with experiments
in Sec. \ref{Sec:Pattern}.

\subsection{Hole doping}
\label{Sec:Hole doping}

\subsubsection{Without JT coupling}
\label{Sec:dopewithoutJT}

In the following, the results of doped states are presented.
Carrier doping leads to an MI transition.
In the absence of the JT coupling,
the metal near the Mott transition shows some anomalous behavior
in our 2D model.

\begin{figure}
\epsfxsize=8cm
\centerline{\epsfbox{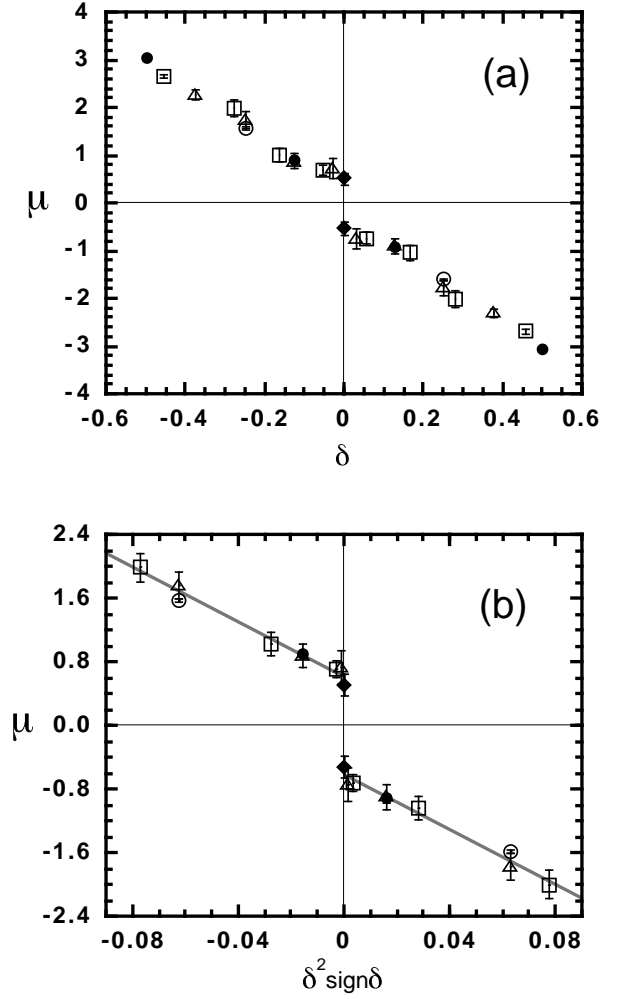}}
\caption{
Doping dependence of the chemical potential for $\tilde{U} = 5$.
The open circles, squares, and triangles correspond to
the data for $L = 4, 6$, and $8$, respectively.
The solid circles are the data for $L = 4$
with the periodic boundary condition.
The solid diamonds at $\delta=0$ show the charge gap amplitude
obtained in Fig. \protect{\ref{Fig:GapwithoutJT}}.
The data are fitted by $\delta^{2}$ in (b).
}
\label{Fig:muvsdelta}
\end{figure}

The first anomaly appears in the charge compressibility.
Figure \ref{Fig:muvsdelta} shows the doping-concentration
dependence of the chemical potential $\mu$.
Here, in order to reduce the finite-size effects,
$\mu$ is calculated from the energy difference
between adjacent closed-shell configurations.
\cite{Furukawa1992}
The chemical potential $\mu$ has a jump at $\delta=0$
due to the finite Mott gap and
changes continuously in the doped region.
The monotonic decrease of $\mu$ with increasing $\delta$ suggests that
there is no phase separation in our 2D model without the JT coupling.
As shown in Fig. \ref{Fig:muvsdelta}(b),
the chemical potential scales to $\delta^{2}$
near the MI transition at $\delta=0$.
This indicates that the compressibility $\kappa \equiv dn/d\mu$
diverges as 
\begin{equation}
        \label{kappascaling}
        \kappa \propto \delta^{-1}
\end{equation}
toward the transition.

\begin{figure}
\epsfxsize=8cm
\centerline{\epsfbox{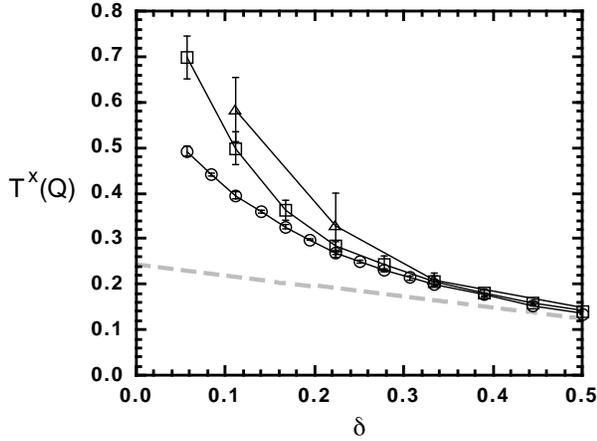}}
\caption{
Doping dependence of the peak values of the orbital correlation function
for $N_{\rm S} = 6\times 6$.
The circles, squares, and triangles correspond to
the data for $\tilde{U} = 3, 4$, and $5$, respectively.
The dotted line shows the result for the noninteracting case.
The lines connecting the data are guides to eye.
}
\label{Fig:TxQUdep}
\end{figure}

Another anomaly is the critical enhancement of
the orbital short-ranged correlations.
In the doped region, the orbital correlation is enhanced
by the Coulomb interaction.
Figure \ref{Fig:TxQUdep} shows the doping dependence
of the peak values of the orbital correlation function defined in Eq. 
(\ref{Tkdef})
for various values of the interaction $\tilde{U}$.
For finite values of $\tilde{U}$,
the correlation shows a diverging behavior
towards the undoped state where the orbital correlation is
long-ranged.
This behavior becomes more conspicuous
with increasing the value of $\tilde{U}$.

\begin{figure}
\epsfxsize=8cm
\centerline{\epsfbox{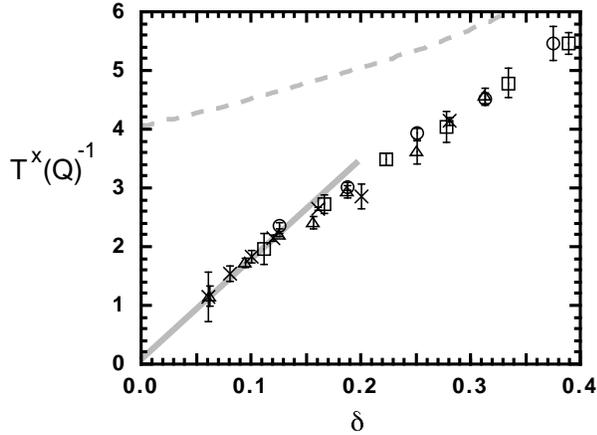}}
\caption{
Doping dependence of the inverse peak values of
the orbital correlation function for $\tilde{U} = 4$.
The circles, squares, triangles, and crosses correspond to
the data for $L = 4, 6, 8$, and $10$, respectively.
The gray line shows the fit by a $\delta$-linear function
for $0<\delta<0.15$.
The dotted line shows the result for the noninteracting case.
}
\label{Fig:TxQvsdelta}
\end{figure}

Figure \ref{Fig:TxQvsdelta} summarizes the data
on the orbital correlations at $\tilde{U}=4$
for various system sizes.
The inverse peak values are plotted in this figure.
The data are independent of system sizes even near $\delta=0$,
which indicates that the orbital correlations are short-ranged.
As shown in this plot, we find a scaling relation given by
\begin{equation}
        \label{TxQscaling}
        T^{x}(\mbox{\boldmath $Q$}) \propto \delta^{-1}
\end{equation}
in the vicinity of the MI transition.
This suggests that the orbital ordering at $\delta=0$ is
immediately destroyed by hole doping and that
the short-ranged correlation diverges towards the transition
in the doped region.

These anomalous properties are in marked contrast with
the usual MI transition from a metal to a band insulator.
The scaling relations, Eqs. (\ref{kappascaling}) and (\ref{TxQscaling}),
are similar to those in the ordinary single-band Hubbard model
in two dimensions.
\cite{Furukawa1992}
This similarity is discussed in Sec. \ref{Sec:Incoherence}
to characterize the MI transition in the present model.
Our results will be examined to consider the charge incoherence
in experiments.

\subsubsection{With JT coupling}
\label{Sec:dopewithJT}

\begin{figure}
\epsfxsize=8cm
\centerline{\epsfbox{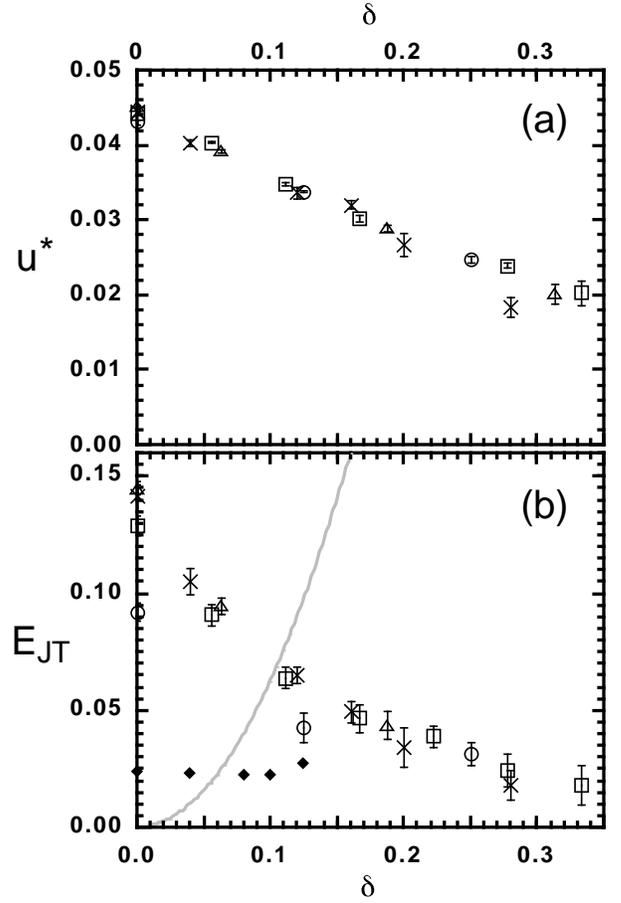}}
\caption{
The JT distortion in the hole-doped region
for $\tilde{U}=4$, $g=10$, and $k=100$;
(a) the oxygen distortion and (b) the stabilization energy.
The circles, squares, triangles, and crosses correspond to
the data for $L = 4, 6, 8$, and $10$, respectively.
The filled diamonds in (b) are the magnetic transition temperature
in experiments.
The gray curve shows the estimation of the kinetic-energy gain
in the $z$ direction given by Eq. ~(\protect{\ref{deltaEtz}}).
}
\label{Fig:uEJTvsdelta}
\end{figure}

Here, the effects of the JT coupling in hole-doped region
are investigated.
Phonons are treated by the mean-field
approximation as in Sec. \ref{Sec:HFwithJT}.
The mean-field treatment generally overestimates
the JT distortion especially in doped cases
since it neglects various fluctuations.
Actually, Fig. \ref{Fig:uEJTvsdelta}(a) shows that
the cooperative JT distortion remains finite
in a wide range of doping.

However the stabilization energy of the JT distortion
decreases rapidly as a function of the doping concentration $\delta$.
Figure \ref{Fig:uEJTvsdelta}(b) shows $\delta$ dependence
of the stabilization energy of the JT distortion.
The rough estimates of the cutoff energy are also shown in the figure,
below which some other energy scales not considered in our model
come into play as discussed in Sec. \ref{Sec:what parameter};
one is the magnetic transition temperature
$T_{\rm N}$ for La$_{1-x}$Sr$_{x}$MnO$_{3}$,
\cite{Kawano1996b}
and the other is the effect of 3D hoppings
$\Delta E_{t}^{z}$ given by Eq. (\ref{deltaEtz}).
In Sec. \ref{Sec:Collapse},
the collapse of the JT distortion
will be discussed with this rapid decrease of $E_{\rm JT}$.

\begin{figure}
\epsfxsize=8cm
\centerline{\epsfbox{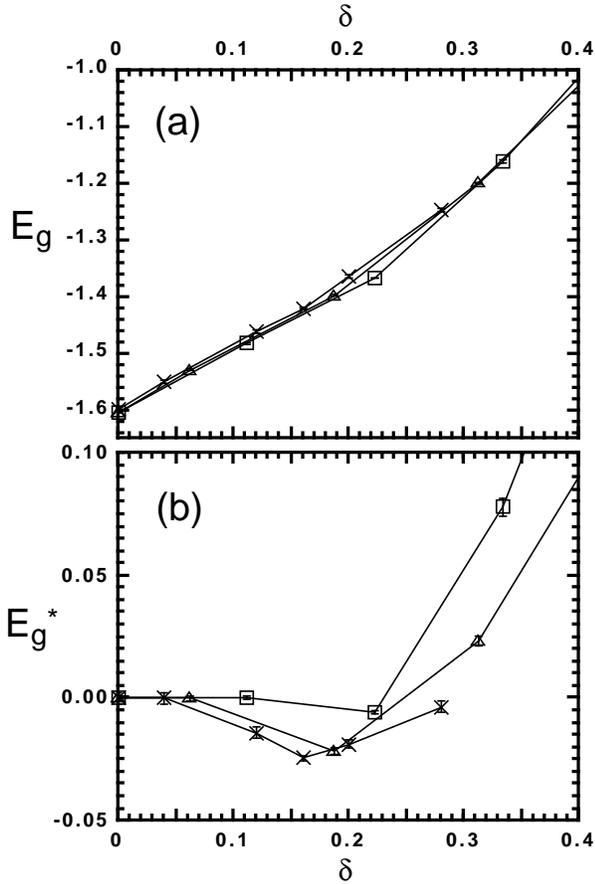}}
\caption{
Doping dependence of the ground state energy per site
for $\tilde{U}=4$, $g=10$, and $k=100$.
The squares, triangles, and crosses correspond to
the data for $L = 6, 8$, and $10$, respectively.
The lines are guides to eye.
In (b), the quantity defined in Eq. ~(\protect{\ref{Eg*def}}) is plotted
to show the convex region clearly. See the text for details.
}
\label{Fig:EgvsdeltaJT}
\end{figure}

Figure \ref{Fig:EgvsdeltaJT}(a) shows doping dependence of
the ground state energy per site $E_{\rm g}$ at $\tilde{U}=4$
in the presence of the JT coupling.
The data show a convex region close to half filling.

In order to see this behavior more clearly,
we consider the quantity in each system size defined by
\begin{equation}
	\label{Eg*def}
	E_{\rm g}^{*} = E_{\rm g} - E_{\rm g}(\delta=0)
	          - \Delta_{\rm c}^{*} \delta,
\end{equation}
where $\Delta_{\rm c}^{*}$ is the slope of $E_{\rm g}$ at $\delta=0$
in Fig. \ref{Fig:EgvsdeltaJT}(a) and represents
the charge gap amplitude obtained as the chemical potential
in the closed-shell adjacent to half filling.
Figure \ref{Fig:EgvsdeltaJT}(b) plots doping dependence of $E_{\rm g}^{*}$.
The convex behavior is clearly seen near the half-filled state.
It becomes more conspicuous with increasing the system size.

The convex region of $E_{\rm g}$ as a function of doping indicates
an instability to phase separation.
The region of the phase separation is determined by
a tangent drawn from the point at $\delta=0$ to the curve of $E_{\rm g}$
in Fig. \ref{Fig:EgvsdeltaJT}(a) or equivalently $E_{\rm g}^{*}$
in Fig. \ref{Fig:EgvsdeltaJT}(b).
States which have a higher energy than this line in the convex region
are unstable and replaced by ones on the line with phase mixing.
The phase mixture consists of the half-filled state and
a doped state at which the tangent line comes
in contact with the energy curve.

In the present model, the phase separation takes place
between the half-filled state with the cooperative JT distortion and
the doped metallic state,
although it is difficult to determine the critical doping concentration
in this case of $\tilde{U}=4$.
Within the mean-field treatment, the latter doped state is
still accompanied by the JT distortion as shown in Fig. 
\ref{Fig:uEJTvsdelta}(a).
In Sec. \ref{Sec:PS}, these results will be examined in detail,
especially to understand the charge-ordered state in experiments.
They will also be compared with other theoretical studies.

\section{Discussion}
\label{Sec:Discussion}

In this section, we discuss our numerical results
shown in the previous section.
For the half-filled states, 
the synergetic effects between the Coulomb interaction and the JT coupling
are clarified in Sec. \ref{Sec:Hierarchy}.
Our results are compared with experiments on perovskite Mn oxides
as well as with other theoretical investigations.
The pattern of orbital ordering is examined in Sec. \ref{Sec:Pattern}.
In Sec. \ref{Sec:metal},
numerical results in the hole-doped states are discussed
in comparison with the recent experiments.
The critical properties of the MI transition,
especially effects of orbital fluctuations, are examined
in Sec. \ref{Sec:Incoherence}
as origins of the experimentally-observed charge incoherence.
Doping effects in the presence of the JT distortions
are also discussed to understand experiments in lightly doped region.
We consider the collapse of the JT distortions
in Sec. \ref{Sec:Collapse} and
the phase separation in Sec. \ref{Sec:PS}.

\subsection{Undoped state}
\label{Sec:Mott}

\subsubsection{Synergy of Coulomb interaction and JT coupling}
\label{Sec:Hierarchy}

Consider first the case that we take account of
only the Coulomb interaction and neglect the JT coupling.
As shown in Sec. \ref{Sec:HFwithoutJT},
the system becomes Mott insulating
with the staggered orbital ordering.
The charge gap opens much slower than the mean-field results,
which shows the mean-field approximation is not appropriate
due to strong quantum fluctuations
in the region of $\tilde{U}$ considered here.
We speculate that the somewhat complicated nesting property
discussed in Appendix B may be related to this strong fluctuation.

This strong fluctuation effect is also inferred from
in the previous studies.
Several mean-field analyses predicted energetic degeneracy
between different spin and orbital structures.
\cite{Kugel1973,Shiina1997,Feiner1999,Maezono1998,Mizokawa1995,Mizokawa1996}
Larger values of $\tilde{U}$ are required
than that in our estimates ($\tilde{U} \simeq 4-5$)
to reproduce experimental results
because they have not taken into account
large contribution from the JT coupling.
Note also that in LDA and LSDA calculations
which generally tend to underestimate
the electron correlation effects,
a ferromagnetic metal would be realized
if the JT distortion is absent.
\cite{Satpathy1996,Solovyev1996a}

On the other hand, if only the JT coupling is taken into account,
we obtain a cooperative JT distortion with an orbital ordering;
however, as shown in Fig. \ref{Fig:GapEJTuvsU},
the JT distortion is too small to reproduce
the realistic oxygen distortions and the charge gap amplitude.
This shortcoming may correspond to
the insufficient charge-gap amplitude predicted
in LSDA using the experimental JT distortions.
\cite{Satpathy1996,Solovyev1996a}

When the effects of both the Coulomb interaction and
the JT coupling are considered as in Sec. \ref{Sec:HFwithJT},
the experimental results on charge, orbital, and lattice degrees of freedom
are quantitatively reproduced within the present model.
Our finding of the synergetic effect by the electron correlation
and the JT coupling suggests that the value of $\tilde{U}$
around $4$ or $5$ would be
sufficient to reproduce experimental results.
Our results give a consistent picture of 
the energetic hierarchy in the undoped state:
(i) The charge gap of the order of $1$eV opens
as the consequence of the synergy between 
the on-site Coulomb interaction and the atomic JT coupling.
These two elements work cooperatively and nonlinearly.
(ii) The correlations of the orbital and the lattice distortion
grow in the energy scale one order of magnitude
smaller than the charge gap amplitude.
The transition temperature of these orderings is
around $0.1$eV.
(iii) The two elements which are neglected in our model, that is,
the three dimensionality and spin degrees of freedom,
may not be relevant to understand the experimental results considered here.
The 3D antiferromagnetic state of $A$-type
appears at the energy much smaller than the above two energy scales,
namely, at the order of $0.01$eV. 
Orbital and lattice states may not be much affected
by the magnetic ordering.

In our analyses, the Coulomb interaction is considered in an
unbiased way in PQMC, while the lattice distortions are treated 
in the mean-field scheme.
Therefore, the effects of the interaction on the orbital 
correlations are fully taken into account
up to the short-ranged and dynamical fluctuations, whereas
such fluctuations of the JT distortions are neglected.
This will lead to the following consequence:
The charge gap amplitude, $\Delta_{\rm c}$, could
be overestimated since the formation of a polaron is not considered.
A doped hole in the insulator may cause a relaxation of the JT 
distortion around it to form a polaron.
Therefore, the charge gap may decrease
due to the formation energy of the polaron.
For the precise estimate of $\Delta_{\rm c}$ under this effect,
it is necessary to fully take into account of the local JT distortions.

Another point to be discussed in our analyses is
the comparison between the stabilization energy of the JT distortion 
$E_{\rm JT}$ and the transition temperature $T_{\rm c}$
of the orbital and lattice orderings.
To be precise, $E_{\rm JT}$ is the energy difference
between the JT-distorted state and the undistorted one
with the orbital ordering stabilized by the Coulomb interaction.
It contains not only the intersite-correlation energy
of the orbital and lattice orderings
but also the on-site JT polarization energy.
On the other hand, the transition temperature $T_{\rm c}$
should be basically determined by the intersite correlations 
of the orbital and lattice orderings
\cite{Kanamori1960}
because the on-site JT polarization may already be
more or less stabilized well above $T_{\rm c}$.
Therefore, in our results,
$E_{\rm JT}$ may overestimate the energy scale of $T_{\rm c}$.

The above two points presumably require larger values of parameters,
$\tilde{U}$, $g$, or $k$ than in the present work.
We need further studies on the JT fluctuation effects and
also on a more appropriate choice of the parameter values.

\subsubsection{Pattern of orbital ordering}
\label{Sec:Pattern}

\begin{figure}
\epsfxsize=6cm
\centerline{\epsfbox{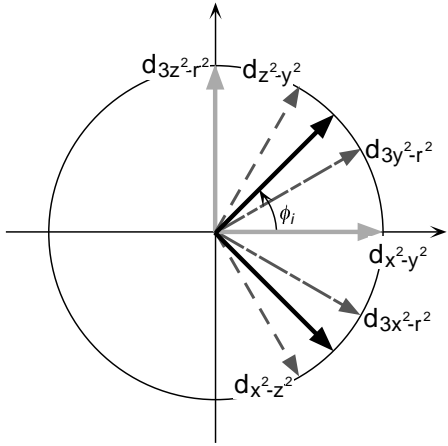}}
\caption{
Single-site orbital state defined as
Eq. ~(\protect{\ref{lc of orbital}}).
The gray arrows at $\phi_{i} = 0$ and $\pi/2$ show
the original basis in the Hamiltonian, $d_{x^{2}-y^{2}}$ and
$d_{3z^{2}-r^{2}}$, respectively.
The bold black arrows at $\phi_{i} = \pm \pi/4$ correspond to
the pattern of the orbital ordering induced by the Coulomb interaction
in our calculations.
The other arrows show the basis at $\phi_{i} = \pm \pi/6$ and $\pm \pi/3$.
See the text for details.
}
\label{Fig:phi}
\end{figure}

First, consider the pattern of the orbital ordering
determined only by the Coulomb interaction.
As described in Sec. \ref{Sec:HFwithoutJT},
our 2D model shows the pattern with 
$\phi_{i} = (-1)^{|\mbox{\boldmath $r$}_{i}^{x} + 
\mbox{\boldmath $r$}_{i}^{y}|} \pi/4$
in terms of Eq. (\ref{lc of orbital}),
which is slightly different from the previous mean-field results
in 3D models as shown in Fig. \ref{Fig:phi}.
In order to understand this discrepancy,
we discuss here effects of the three dimensionality
neglected in our 2D model.
In the $z$ direction, the hopping integral has a strong orbital dependence;
whose explicit form is given by
$\tilde{t}_{ij}^{11}=\tilde{t}_{ij}^{12}=\tilde{t}_{ij}^{21}=0$ and
$\tilde{t}_{ij}^{22}=-\tilde{t}_{0}$.
This favors larger occupation of
the $d_{3z^{2}-r^{2}}$ orbital in 3D systems than in 2D
since only the $d_{3z^{2}-r^{2}}$ orbital
has an overlap to the $z$ direction.
Therefore, we expect that the orbital-ordering pattern
obtained in the present work, that is,
$( |d_{x^{2}-y^{2}}\rangle + |d_{3z^{2}-r^{2}}\rangle) /
( |d_{x^{2}-y^{2}}\rangle - |d_{3z^{2}-r^{2}}\rangle)$ type
may be modified
in a way that a uniform component of $d_{3z^{2}-r^{2}}$ increases.
In other words, our result $|\phi_{i}| = \pi/4$ gives
a lower bound for the angle $|\phi_{i}|$ in the 3D models.
The resultant ordering may look like
$|d_{z^{2}-x^{2}}\rangle / |d_{y^{2}-z^{2}}\rangle$ type.
See Fig. \ref{Fig:phi}.
Actually, this type of orbital ordering was obtained
by the mean-field approximation for 3D electron systems.
\cite{Kugel1973,Shiina1997,Feiner1999,Maezono1998,Mizokawa1995,Mizokawa1996}

When the JT coupling is taken into account,
as shown in Fig. \ref{Fig:anglephi},
the lattice distortion modifies the orbital ordering.
Our result at $\tilde{U} = 4$ gives $|\phi_{i}| \simeq \pi/5$,
which is very close to
$| d_{3x^{2}-r^{2}}\rangle$/$| d_{3y^{2}-r^{2}}\rangle$-type.
Since our calculations neglect the three dimensionality,
the angle $|\phi_{i}| \simeq \pi/5$ may give a lower bound
as discussed above.

Recently, the staggered orbital ordering in LaMnO$_{3}$
was experimentally observed
by the resonant X-ray scattering technique,
\cite{Murakami1998}
although this experiment has not succeeded in determining
the angle $|\phi_{i}|$ unfortunately.
In principle, the angle $|\phi_{i}|$ can be determined
from the intensity of the fundamental reflection in this experiment.
\cite{Ishihara1998}
More detailed experiments are desired to compare our numerical
results.

\subsection{Doped state}
\label{Sec:metal}

\subsubsection{Incoherence in charge dynamics}
\label{Sec:Incoherence}

Here, the critical properties of the MI transition
in our 2D model  are analyzed in the absence of the JT coupling.
As mentioned in Sec. \ref{Sec:dopewithoutJT},
the metal shows two anomalous properties near the MI transition;
the critical divergence of both the compressibility and
the orbital short-ranged correlation.
In the 2D ordinary Hubbard model, the same scaling relations
as Eqs. (\ref{kappascaling}) and (\ref{TxQscaling})
have been reported in detailed numerical investigations.
\cite{Furukawa1992}
The scaling theory based on the hyperscaling hypothesis describes
this MI transition with the novel universality class
characterized by $z=1/\nu=4$,
\cite{Imada1995}
where $z$ is the dynamical exponent and $\nu$ is
the correlation-length exponent.
Recently, other anomalies in this MI transition have also been found in
the localization length in the insulating state
\cite{Assaad1996b}
and the flat dispersion near $\mbox{\boldmath $k$} = (\pi,0)$.
\cite{ImadaPREPRINT}
These two also support the scaling statement.
The scaling relations in our model,
Eqs. (\ref{kappascaling}) and (\ref{TxQscaling}),
imply that the MI transition in the present model
is in a similar universality class to the ordinary Hubbard model.

This argument may give a clue to understand
the strong charge incoherence near the MI transition
observed in perovskite Mn oxides.
\cite{Okimoto1995,Okimoto1997}
In the ordinary Hubbard and $t$-$J$ models,
the optical conductivity has been numerically investigated
to clarify the strong incoherence in charge dynamics.
\cite{NakanoUNPUBLISHEDa,Tsunetsugu1998}
In these models, the residual entropy and
the spin short-ranged correlations are enhanced
towards the MI transition point, which may be the origin of
such charge incoherence.
The role of spins in these models is replaced with
orbital degrees of freedom in the present model (\ref{H el}).
This suggests that the orbital correlations induced by the Coulomb 
interaction may be one of the driving mechanisms of
the experimentally-observed charge incoherence.
Actually, we have also calculated the optical conductivity
in the present model by the exact diagonalization of small clusters
and obtained strongly incoherent properties as expected.
These results will be reported elsewhere in detail.
\cite{NakanoUNPUBLISHEDb}

Experimentally, the incoherent charge dynamics is observed
not only in the 3D materials but also in the 2D ones such as
La$_{2-2x}$Sr$_{1+2x}$Mn$_{2}$O$_{7}$ near $x = 0.4$.
\cite{Ishikawa1998}
This fact suggests that the essence of the charge incoherence
in these materials is captured by the models of metals
near the MI transition from strongly 2D anisotropic insulators.
The scaling argument assuming such an anisotropy in energy dispersions
has been discussed recently by one of the authors.
\cite{Imada1998a}
It implies that the system has a 2D anomalous dispersion
with a flat $\mbox{\boldmath $k$}$-dependence caused
by the critically-enhanced 2D orbital correlations.
This argument provides one possible scenario
to reconcile the puzzling situations in the 3D materials
which were mentioned in Sec. \ref{Sec:Introduction};
the specific-heat coefficient $\gamma$ remains small
and is not critically enhanced
\cite{Woodfield1997}
while the Drude weight $D$ and the discontinuity at the Fermi edge
are strongly suppressed in the critical region.
\cite{Okimoto1995,Okimoto1997,Sarma1996}
Our results support the existence of the 2D anomalous dispersion
in the charge excitation near the Mott transition
in the present model.
To confirm this scenario further,
it is necessary to investigate the $\mbox{\boldmath $k$}$-dependence
of the dispersion directly.
This is left for future study.

\subsubsection{Collapse of JT distortion}
\label{Sec:Collapse}

In real materials, the cooperative JT distortions persist
up to finite doping concentration in the spin-canted state,
for instance, up to $\delta \simeq 0.1$
in La$_{1-x}$Sr$_{x}$MnO$_{3}$.
\cite{Kawano1996a,Kawano1996b}
In our results in Sec. \ref{Sec:dopewithJT},
although the JT distortion loses its stability rapidly
with increasing the doping concentration,
it remains finite in a wide range of doping.
This presumably results from the mean-field treatment in which
the fluctuations of the JT distortions are neglected.
The doped holes may form polarons, which leads to
an importance of the so-called breathing mode of phonons.
\cite{Millis1996b}
Therefore, the short-ranged JT fluctuations should be
taken into account more seriously
in the doped region than in the undoped state.
To precisely estimate the doping concentration
where the JT distortion collapses,
it is necessary to fully consider the JT fluctuations.

Since the distortion has strong 2D anisotropy,
the three dimensionality may also be important to understand
the collapse of the JT distortion.
In Fig. \ref{Fig:uEJTvsdelta}(b),
the energy scales discussed
in Sec. \ref{Sec:what parameter} are plotted for comparison;
the magnetic transition temperature $T_{\rm N}$ and
the kinetic-energy gain by the three dimensionality,
$\Delta E_{t}^{z}$.
Our model may not be appropriate to discuss phenomena
happening in the energy scale comparable to them.
In order to consider the JT distortion for $\delta \simge 0.1$
where $E_{\rm JT}$ becomes comparable to $\Delta E_{t}^{z}$,
we may need to include 3D effects.

The above two points require further studies on
doping effects with the JT distortions.
The former, the fluctuation effects of the JT distortions,
may be partly taken into account within the present model
(\ref{H tot}) if we fully optimize the parameters
$u_{i}$ and $\theta_{i}$.

\subsubsection{Phase separation}
\label{Sec:PS}

Our results in the doped region with the JT coupling
indicate the existence of the phase separation.
This phase separation is a consequence of two factors;
one is the critically-enhanced compressibility
which exists even in the absence of the JT coupling
as in Eq. (\ref{kappascaling}), and
the other is the large stabilization energy $E_{\rm JT}$
which decreases rapidly with increasing doping.
Therefore, even if the JT fluctuation effects are taken into account,
the phase separation may persist in the present model.

In real systems, the additional elements such as the long-ranged
Coulomb interaction might be crucial to replace
the phase separation by another instability, for instance,
an ordering of the hole distribution.
Recently, the charge ordering has been reported
in La$_{1-x}$Sr$_{x}$MnO$_{3}$ around $x = 0.125$.
\cite{Yamada1996,Zhou1997}
From neutron-scattering data,
this phase has been interpreted as
a stacking of two alternating layers;
one has little hole with a similar JT distortion
to the undoped state, and
the other contains excess holes and
shows no apparent JT distortion.
This phase is favorably compared to our result of the phase separation
because it predicts the phase mixture of
the JT-distorted undoped state and the doped one.
The latter doped phase may show no apparent JT distortion
if we take account of the JT fluctuations
as mentioned in Sec. \ref{Sec:Collapse}.

Recently, the phase separation has also been discussed
in the original `simple' DE model given by Eq. (\ref{H DE}).
\cite{Yunoki1998a,Dagotto1998,Yunoki1998b}
In these studies,
the system shows the phase separation between
an undoped antiferromagnetic insulator and
a doped ferromagnetic metal at low temperatures.
Effects of the JT distortion on this phase separation
have also been discussed.
\cite{YunokiPREPRINT}
In these cases, the phase separation originates from
a change in the density of states between two states;
the antiferromagnetic insulator stabilized
by superexchange interaction has a narrow band
while the ferromagnetic metal dominated by the DE interaction
has a rather wide band.
Note that in the `simple' DE model, the phase separation takes place
even in the absence of the Coulomb interaction
between the conduction electrons.
On the other hand, in our model, the Coulomb interaction plays
a crucial role to induce the phase separation as mentioned above.

\subsubsection{Remark}
\label{Sec:Remark}

Our numerical results may give a comprehensive scenario
to understand the complicated properties
in lightly-doped Mn oxides:
Doped holes induce the JT fluctuations.
The spin canting enhances the 3D motions of doped holes.
These may contribute cooperatively
to collapse the cooperative JT distortion
which becomes unstable rapidly at a small doping concentration.
On the other hand, the critical region of the MI transition
may be wider than the critical concentration of the collapse.
Note that the critical region of the Mott transition spreads
up to $\delta \simeq 0.2-0.3$ in the case of $\tilde{U}=5$
as shown in Fig. \ref{Fig:muvsdelta}(b).
Then, the doped state beyond the collapse of the JT distortion
shows the strong incoherence in charge dynamics.
In these lightly-doped regions, charge degrees of freedom
have an instability suggested by the phase separation in our results.
At a commensurate filling, for instance, at $\delta=0.125=1/8$,
doped holes may form a hyperlattice structure
as indicated in experiments.

\section{Summary}
\label{Sec:Summary}

In this work, roles of orbital and lattice degrees of freedom
in the extended double-exchange models
have been investigated to understand remarkable properties of
undoped and lightly-doped perovskite Mn oxides.
The models contain the Coulomb interaction, double degeneracy
of the conduction bands and the Jahn-Teller coupling.
The ground state properties are studied
under full polarization of spins with two-dimensional anisotropy
by using the quantum Monte Carlo method.
Fluctuation effects of Coulomb interaction and
orbital degrees of freedom are investigated
without biased approximations.

In the undoped states, the importance of
both the Coulomb interaction and the Jahn-Teller coupling is clarified.
In the absence of the Jahn-Teller coupling,
the system shows strong quantum fluctuations
with remarkable suppression of the Mott gap amplitude.
A realistic amplitude of the Jahn-Teller coupling suppresses
the fluctuations and stabilizes the insulating state.
Synergy between the Coulomb interaction and the Jahn-Teller coupling
remarkably enhances the charge gap amplitude
and reproduces the realistic Jahn-Teller distortion.
Our numerical results give us a consistent picture of
the undoped states, especially of the characteristic hierarchy
of energy scales among charge, orbital-lattice, and spin
degrees of freedom.
The energy hierarchy consists of the charge gap $\sim 1$eV,
orbital and lattice ordering energy $\sim 0.1$eV, and
spin ordering and 3D coupling scales $\sim 0.01$eV.
The pattern of staggered orbital orderings has also been
investigated in detail.
It would be desired to compare it
with more detailed experiments in the future.

Upon doping of carriers,
the critical properties of the metal-insulator transition
have been investigated
in the absence of the Jahn-Teller distortion.
Our numerical results show two anomalies;
the critical enhancement of the compressibility and
the orbital short-ranged correlations towards the transition.
These results are analyzed by using the scaling argument
in comparison with the ordinary Hubbard model.
The orbital fluctuation enhanced by the Coulomb interaction
is discussed as one of the driving mechanisms of 
the strong incoherence in charge dynamics observed in experiments.

Next, the effects of the Jahn-Teller coupling
in the hole-doped region have been investigated.
Our mean-field treatment of phonons suggests that
the cooperative Jahn-Teller distortion becomes unstable
rapidly with increasing hole doping.
Moreover, the instability of phase separation appears
near the undoped state.
These results are favorably compared with
the collapse of the Jahn-Teller distortion and
the charge-ordered phase observed in experiments.
We propose a scenario to understand the rich physics
of the lightly-doped Mn oxides;
the order-disorder Jahn-Teller transition,
the charge ordering near
the magnetic and metal-insulator transition, and
the strong charge incoherence.
More detailed study on fluctuation effects of
the Jahn-Teller distortions as well as on the three dimensionality
is left for further study.

\section*{Acknowledgement}

The authors thank Junjiro Kanamori and Hiroki Nakano
for fruitful discussions.
A part of the computations in this work was performed
by using the facilities of the Supercomputer Center,
Institute of Solid State Physics, University of Tokyo.
This work is supported by `Research for the Future Program'
from Japan Society for Promotion of Science (JSPS-RFTF 97P01103).
Y. M. acknowledges the financial support of
Research Fellowships of Japan Society for the Promotion
of Science for Young Scientists.

\appendix

\section{}

In this Appendix, we discuss the negative sign problem in PQMC
for the model given by Eq. (\ref{H el}).
The ratio of the negative sign in QMC sampling depends on
the system size $N_{\rm S}$, the Coulomb interaction $\tilde{U}$,
the trial wave function, and the doping concentration $\delta$.
\cite{Loh1990,Furukawa1991b}

First, it should be pointed out that
PQMC for the present model shows negative-sign samples
even at the particle-hole symmetric point,
that is, at half filling ($\mu=0$).
In contrast to this, for the ordinary Hubbard model,
the negative sign problem completely disappears at half filling;
\cite{Hirsch1983,Hirsch1985}
the system is mapped to an attractive Hubbard model
under the transformations as
\begin{equation}
        \label{phsym1}
        c_{i\uparrow}^{\dagger} \rightarrow (-1)^{i} 
        \tilde{c}_{i\uparrow}, \quad
        c_{i\downarrow}^{\dagger} \rightarrow 
        \tilde{c}_{i\downarrow}^{\dagger},
\end{equation}
which ensure the positivity of samples in QMC.
This mapping does not apply to the present model (\ref{H el})
because of the difference in the diagonal hoppings, $t^{11} \ne t^{22}$,
and the off-diagonal elements $t^{12}$, $t^{21}$.
This leads to the negative sign problem even at half filling.

\begin{figure}
\epsfxsize=8cm
\centerline{\epsfbox{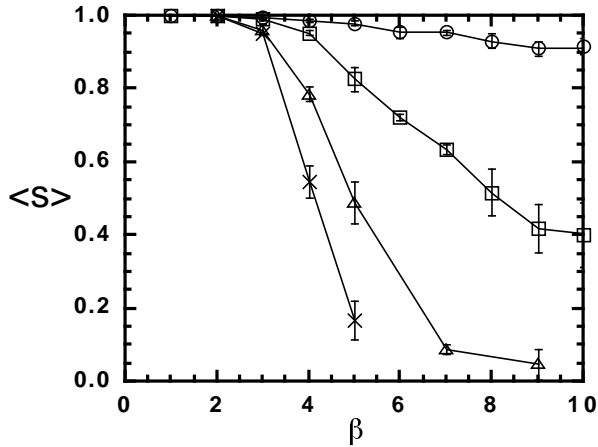}}
\caption{
$\beta$ dependence of the negative sign for $\tilde{U} = 4$
and $g=0$ at half filling.
The circles, squares, triangles, and crosses correspond to
the data for $L = 4, 6, 8$, and $10$, respectively.
The orbital-singlet state is used as the trial wave function.
The lines are guides to eye.
}
\label{Fig:NSdepofS}
\end{figure}

\begin{figure}
\epsfxsize=8cm
\centerline{\epsfbox{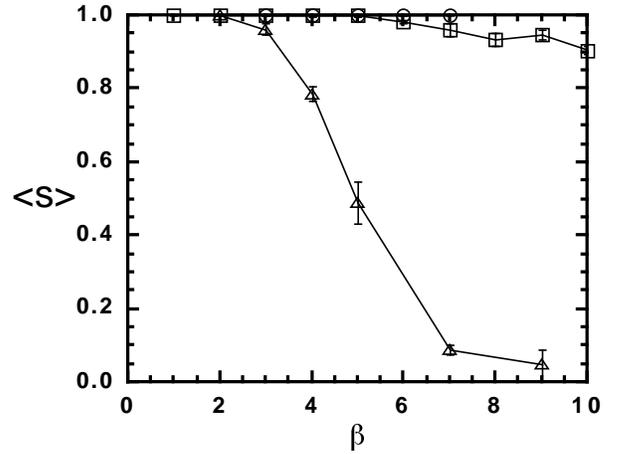}}
\caption{
$\beta$ dependence of the negative sign for $N_{\rm S} = 8\times 8$
at half filling with $g=0$.
The circles, squares, and triangles correspond to
the data for $\tilde{U} = 2, 3$, and $4$, respectively.
The orbital-singlet state is used as the trial wave function.
The lines are guides to eye.
}
\label{Fig:UdepofS}
\end{figure}

\begin{figure}
\epsfxsize=8cm
\centerline{\epsfbox{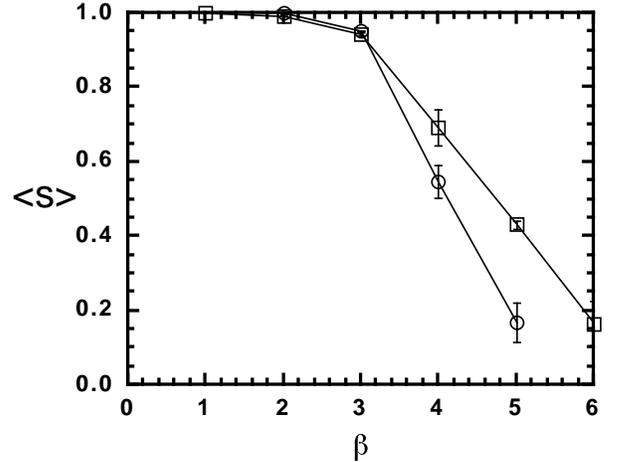}}
\caption{
$\beta$ dependence of the negative sign for $\tilde{U} = 4$
and $g=0$
in the system with $N_{\rm S} = 10\times 10$ at half filling.
The circles and squares correspond to
the data on the orbital-singlet wave function
and on the unrestricted Hatree-Fock solution with $\tilde{U} = 2.1$, respectively.
The lines are guides to eye.
}
\label{Fig:phidepofS}
\end{figure}

Figures \ref{Fig:NSdepofS}-\ref{Fig:phidepofS} show
$\beta$ dependence of the expectation value of
the sign of the MC samples, $\langle S \rangle$, at half filling.
In the figures, $\beta$ is the projection parameter;
the ground state is projected out
by the operation of $\exp(-\beta {\cal H})$ on a trial wave function.
As shown in Figs. \ref{Fig:NSdepofS} and \ref{Fig:UdepofS},
larger system size or larger interaction makes
the negative sign problem more serious.

\begin{figure}
\epsfxsize=8cm
\centerline{\epsfbox{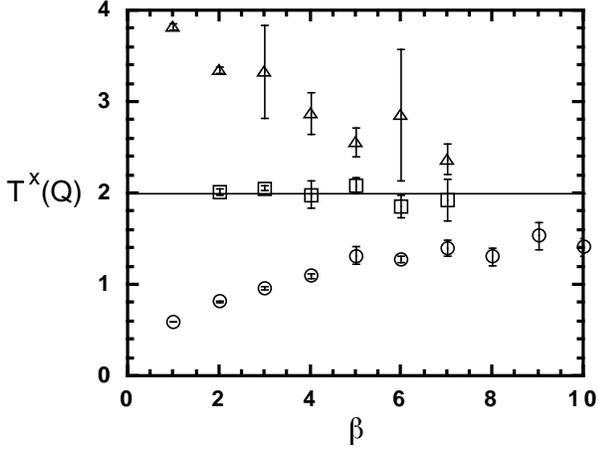}}
\caption{
Convergence for $\beta$ of the orbital correlation function
at the peak point for $\tilde{U} = 4$ and $g=0$
in the system with $N_{\rm S} = 6\times 6$ at half filling.
The circles, squares, and triangles correspond to
the data obtained from the orbital-singlet wave function,
the unrestricted Hartree-Fock solution with $\tilde{U} = 2.4$ and $3$,
respectively.
The horizontal line shows the converged value.
}
\label{Fig:TxQvsbeta}
\end{figure}

Figure \ref{Fig:phidepofS} indicates that it is important
to choose an appropriate trial state in decreasing the sign problem.
In the present work, the half-filled cases are studied
by using the unrestricted Hartree-Fock solutions and
the noninteracting Slater determinant or
the orbital-singlet wave functions
depending on the values of $\tilde{U}$.
The unrestricted Hartree-Fock solutions are optimized
with the value of the interaction $\tilde{U}$
used in the Hartree-Fock calculations
to give the best convergence
as shown in Fig. \ref{Fig:TxQvsbeta} later.
\cite{Furukawa1991a}
The orbital-singlet wave functions are given by the system
with a tiny dimerization introduced in the hopping integrals
to ensure that the ground state is singlet.
\cite{Assaad1997}

\begin{figure}
\epsfxsize=8cm
\centerline{\epsfbox{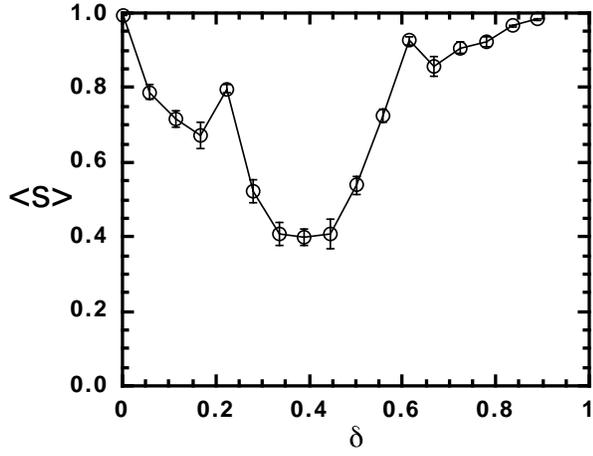}}
\caption{
Doping dependence of the negative sign at $\beta=5$ for $\tilde{U} =
3$ and $g=0$
in the system with $N_{\rm S} = 6\times 6$
for the choice of the orbital-singlet trial functions.
The lines are guides to eye.
}
\label{Fig:deltadepofS}
\end{figure}

Figure \ref{Fig:deltadepofS} shows
$\delta$ dependence of $\langle S \rangle$ at $\beta = 5$.
We note that $\langle S \rangle$ has the largest value at half filling.
The negative sign problem becomes severe
in the lightly doped region, $0 < \delta \simle 0.5$.
For the doped cases, we use the noninteracting Slater determinant
or the orbital-singlet wave functions as the trial state.

The convergence of physical quantities with increasing $\beta$
strongly depends on the choice of trial wave functions.
As an example,
Fig. \ref{Fig:TxQvsbeta} shows the peak value of
the orbital correlation function at half filling.
In this case, a fast convergence is obtained
if we control the input value of $\tilde{U}$
in the unrestricted Hartree-Fock calculations.
\cite{Furukawa1991a}
Generally, the choice of input $\tilde{U}$
somewhat less than the value of the interaction in the model
makes the fastest convergence.
This may be related to the large deviations of the QMC results
from the mean-field ones due to the strong quantum fluctuations
discussed in Sec. \ref{Sec:HFwithoutJT}.

\section{}

Here, the noninteracting case ($\tilde{U}=0$)
in the 2D model (\ref{H el}) is discussed.
We show the energy dispersions of this model and Fermi surfaces at half filling
which have a perfect nesting property.
We also discuss the boundary conditions and
finite-size effects in the linear dimension $L$ of the system.
These considerations are helpful for systematic analyses of
numerical results in Sec. \ref{Sec:Results}.

\begin{figure}
\epsfxsize=5cm
\centerline{\epsfbox{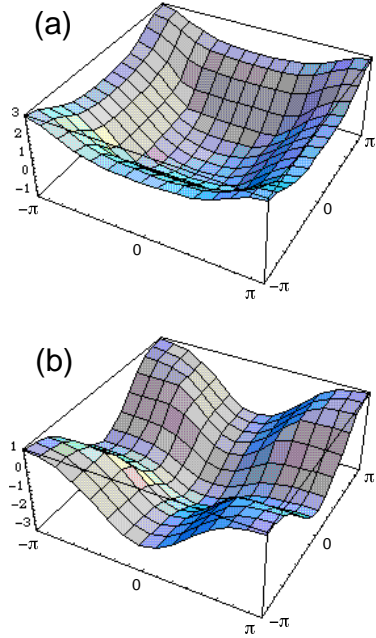}}
\caption{
Energy dispersions for the noninteracting system
defined in Eq. ~(\protect{\ref{epsilonpm}});
(a) $\varepsilon_{+}$ and (b) $\varepsilon_{-}$.
}
\label{Fig:energy disp}
\end{figure}

By Fourier transformation,
the hopping matrix $\tilde{t}_{ij}^{\nu\nu'}$ 
for $u_{i}=0$ in Eq. (\ref{H el}) is denoted as
\begin{equation}
\label{epsilon}
\varepsilon^{\nu\nu'}(\mbox{\boldmath $k$}) =
\varepsilon_{0}(\mbox{\boldmath $k$}) \delta_{\nu\nu'} +
\varepsilon_{1}(\mbox{\boldmath $k$}) \mbox{\boldmath $\tau$}^{z}_{\nu\nu'} +
\varepsilon_{2}(\mbox{\boldmath $k$}) \mbox{\boldmath $\tau$}^{x}_{\nu\nu'},
\end{equation}
where $\delta_{\nu\nu'}$ is the Cronecker's delta function and
$\mbox{\boldmath $\tau$}^{\mu}$ is $\mu$ component of the Pauli matrix.
Here we define
\begin{eqnarray}
\varepsilon_{0} &=& -t_{0} \left( \cos k_{x} + \cos k_{y} \right) \\
\varepsilon_{1} &=& -t_{0}/2 \left( \cos k_{x} + \cos k_{y} \right) \\
\varepsilon_{2} &=& \sqrt{3}t_{0}/2 \left( \cos k_{x} - \cos k_{y} \right).
\end{eqnarray}
The $2\times 2$ matrix $\varepsilon^{\nu\nu'}(\mbox{\boldmath $k$})$ is easily
diagonalized to derive energy dispersions in the form
\begin{equation}
        \label{epsilonpm}
\varepsilon_{\pm} = \varepsilon_{0}
\pm \sqrt{\varepsilon_{1}^{2} + \varepsilon_{2}^{2}},
\end{equation}
which are shown in Fig. \ref{Fig:energy disp}.

\begin{figure}
\epsfxsize=5cm
\centerline{\epsfbox{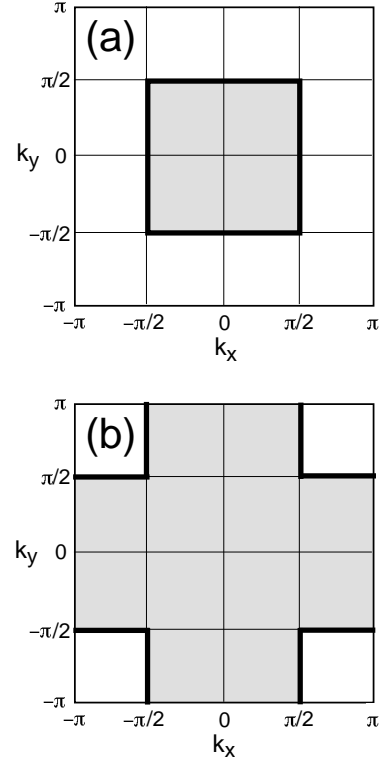}}
\caption{
Fermi surfaces in the noninteracting system at half filling;
(a) for the branch $\varepsilon_{+}$ and (b) for $\varepsilon_{-}$.
The bold solid lines show the Fermi surfaces and the shaded regions
inside them are the occupied states.
}
\label{Fig:FS}
\end{figure}

At half filling, we obtain the Fermi surfaces shown in Fig. \ref{Fig:FS}.
Note that they satisfy a perfect nesting condition, that is,
$\mbox{\boldmath $k$}(\varepsilon_{+}=0) =
\mbox{\boldmath $k$}(\varepsilon_{-}=0) + \mbox{\boldmath $Q$}$, where $\mbox{\boldmath $Q$} = (\pi,\pi)$.
Therefore, we expect that the system may become an insulator
when we switch on the interaction $\tilde{U}$.

It should be stressed that the Fermi surface in Fig. \ref{Fig:FS}
has many different ways of nesting.
Besides the above-mentioned nesting with the wave vector
$\mbox{\boldmath $Q$}$ between two different branches,
it is possible to nest the Fermi surfaces in the same branch
with wave vectors $(\pi,a)$ or $(a,\pi)$
where $a$ is an arbitrary number between $0$ and $\pi$.
In addition, the nesting between different branches
with $(\pi,a)$ or $(a,\pi)$ is also possible.
Therefore, our 2D model (\ref{H el}) may have
a competition between these various nestings at half filling,
which may lead to the strong quantum fluctuations
as seen in numerical results in Sec. \ref{Sec:HFwithoutJT}.

At half filling, finite-size systems with $N_{\rm S}=L^{2}$ sites
are devided in two different series
depending on a linear dimension $L$ and boundary conditions:
One is a group which has $\mbox{\boldmath $k$}$-points
on the Fermi surfaces and the other does not have.
In other words, the former has gapless
but the latter has finite gap excitations
in noninteracting cases.
The former is a series in which
$L=4n$ ($n$ is an integer) with the periodic boundary conditions or
$L=4n+2$ with the antiperiodic boundary conditions, and
the latter is $L=4n$ with the antiperiodic boundary conditions or
$L=4n+2$ with the periodic boundary conditions.

\begin{figure}
\epsfxsize=8cm
\centerline{\epsfbox{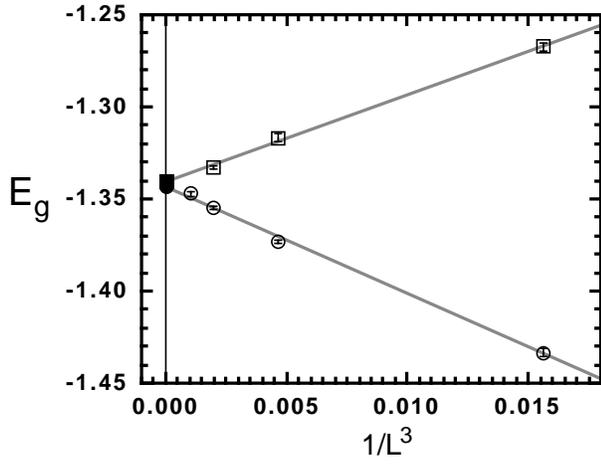}}
\caption{
System-size extrapolation of the ground state energy for $\tilde{U} =
3$ and $g=0$
at half filling.
The circles are the data on the series of $L = 4n$ with
the antiperiodic boundary condition or $L=4n+2$ with
the periodic boundary condition.
The squares are the data on the series of $L = 4n$ with
the periodic boundary condition or $L=4n+2$ with
the antiperiodic boundary condition.
The data are fitted by $1/L^{3}$ and the extrapolated values
to $L \rightarrow \infty$ are shown as the filled symbols.
}
\label{Fig:twoseriesExt}
\end{figure}

These two series should be independently extrapolated
to the thermodynamic limit $N_{\rm S}\rightarrow \infty$.
Figure \ref{Fig:twoseriesExt} shows these extrapolations
of the ground state energy.
The extrapolated values in two series agree well
within the errorbars.
In Sec. \ref{Sec:Results}, 
numerical calculations and extrapolations to the thermodynamic limit
are carefully done by considering these different series separately.

In the doped cases, in order to reduce finite-size effects,
we consider mainly the electron fillings
at which the closed-shell configuration is satisfied
in the noninteracting case.
\cite{Furukawa1992}
As seen in Sec. \ref{Sec:dopewithoutJT} and \ref{Sec:dopewithJT},
especially for the calculations of the chemical potential,
this procedure works well and gives a smooth $\delta$ dependence
insensitive to the system sizes
except for the phase-separated region.


\end{document}